\documentclass[twocolumn]{aastex63}

\usepackage{graphicx, natbib, romannum, verbatim, amsmath}

\accepted{July 28, 2019}

\shorttitle{The quasar LF at $z\sim5$ constructed by DL and BIC}
\shortauthors{Shin, Im $\&$ Kim}

\begin{document}
\title{The quasar luminosity function at $z\sim5$ via deep learning and Bayesian information criterion} 

\author[0000-0002-2188-4832]{Suhyun Shin}
\affil{SNU Astronomy Research Center (SNUARC), Astronomy Program, Department of Physics \& Astronomy, Seoul National University, 1 Gwanak-ro, Gwanak-gu, Seoul 08826, Republic of Korea}
\email{suhyun.shin.s2@gmail.com, myungshin.im@gmail.com}

\author[0000-0002-8537-6714]{Myungshin Im}
\affil{SNU Astronomy Research Center (SNUARC), Astronomy Program, Department of Physics \& Astronomy, Seoul National University, 1 Gwanak-ro, Gwanak-gu, Seoul 08826, Republic of Korea}

\author[0000-0003-1647-3286]{Yongjung Kim}
\affiliation{Department of Astronomy and Atmospheric Sciences, College of Natural Sciences, Kyungpook National University, Daegu 41566, Republic of Korea}
\affiliation{Kavli Institute for Astronomy and Astrophysics, Peking University, Beijing 100871, People's Republic of China}

\begin{abstract}
Understanding the faint end of quasar luminosity function at a high redshift is important since the number density of faint quasars is a critical element in constraining ultraviolet (UV) photon budgets for ionizing the intergalactic medium (IGM) in the early universe. Here, we present quasar LF reaching $M_{1450} \sim -22.0$ AB mag at $z\sim5$, about one magnitude deeper than previous UV LFs. We select quasars at $z\sim5$ with a deep learning technique from deep data taken by the Hyper Suprime-Cam Subaru Strategic Program (HSC-SSP), covering a 15.5 deg$^2$ area. Beyond the traditional color selection method, we improved the quasar selection by training an artificial neural network for distinguishing $z\sim5$ quasars from non-quasar sources based on their colors and adopting the Bayesian information criterion that can further remove high-redshift galaxies from the quasar sample. When applied to a small sample of spectroscopically identified quasars and galaxies, our method is successful in selecting quasars at $\sim83 \%$ efficiency ($5/6$) while minimizing the contamination rate of high-redshift galaxies ($1/8$) by up to three times compared to the selection using color selection alone ($3/8$). The number of our final quasar candidates with $M_{1450} < -22.0$ mag is 35. Our quasar UV LF down to $M_{1450} = -22$ mag or even fainter ($M_{1450} = -21$ mag) suggests a rather low number density of faint quasars and the faint-end slope of $-1.6^{+0.21}_{-0.19}$, favoring a scenario where quasars play a minor role in ionizing the IGM at high redshift.

\end{abstract}
\keywords{cosmology: observations – galaxies: active – galaxies: high-redshift – quasars: supermassive blackholes, methods:data analysis, methods:statistical}

\defcitealias{KimYJ+2020}{K20}

\bigskip
\section{Introduction} \label{sec:intro}
Quasars are the most luminous sub-population of active galactic nuclei (AGNs) powered by the accretion of surrounding mediums to the supermassive black hole located at the center of its host galaxy. Although quasars contribute to maintaining the ionized state of IGM along with star-forming galaxies in the post-reionization era ($z<6$), the role of the quasar in explaining the ionizing background of the universe is not fully understood \citep{Fan+2006, Glikman+2011, Ikeda+2011, Giallongo+2015, Parsa+2018, Boutsia+2018}.

\indent To evaluate the contribution of quasars to the IGM ionizing photon budget, many studies searched for high-redshift quasars \citep{Glikman+2011, Akiyama+2018, Matsuoka+2016, McGreer+2018, Parsa+2018, Giallongo+2019, KimYJ+2015, KimYJ+2019, KimYJ+2020, Wang+2019, Grazian+2020, Shin+2020}, especially at the absolute magnitude at 1450 \AA{} in the rest frame ($M_{1450}$) $\sim -23.5$ mag where the ionizing emissivity of the quasar is considerable, as shown in \citeauthor{KimYJ+2020} (\citeyear{KimYJ+2020}; hereafter \citetalias{KimYJ+2020}). While the quasar UV LFs from different studies are now converging toward a common shape at $M \lesssim -23.5$ mag that can be approximated with a pure number density evolution at $z > 2$ \citep{KimYJ+2021}, there remains great uncertainty in the LF at a fainter magnitude. If the number density of quasars is as high as some studies suggest at the faintest end \citep{Boutsia+2018, Giallongo+2019, Grazian+2020}, quasars are still a viable candidate to be responsible for cosmic re-ionization at high redshift.

\defcitealias{Niida+2020}{N20}

\indent To extend the faint limit of the quasar LF, we can adopt two approaches: one to use deeper data, and another to select quasars using a new technique. As for the deeper data, the second public data release (PDR2; \citealt{Aihara+2019}) of the Hyper Suprime-Cam Subaru Strategic Program (HSC-SSP; \citealt{Aihara+2018}) provides an interesting opportunity. The deeper layers of the survey go down to $i \sim 27.0$ mag \citep{Aihara+2019} and yet an area wide enough (tens of deg$^2$) to negate the cosmic variance in number density. Therefore, such a dataset is suitable for finding quasars with $M_{1450} \sim -22.0$ mag and probing the quasar LF $\sim 1.0$ mag deeper than previously constructed quasar LFs at $z\sim5$ (\citealt{McGreer+2018, Shin+2020, Niida+2020}, hereafter, \citetalias{Niida+2020}). 

\indent Another difficulty in extending the quasar LF to a fainter limit is that galaxies occupy a significant fraction of the high-redshift UV sources at $M_{1450} > -23$ mag (\citealt{Ono+2018, Adams+2020}; \citetalias{Niida+2020}; \citealt{Bowler+2021, KimYJ+2021}) and can contaminate quasar samples made from a conventional color-selection technique. Also, as we explore the fainter limit, the number of sources becomes formidably large. This makes it challenging to apply time-consuming selection methods such as spectral energy distribution (SED) fitting for the quasar selection \citep{Reed+2017}, although such methods may be efficient in discerning quasars from galaxies.

\indent Therefore, we adopt a new and powerful approach that combines deep learning (DL) and Bayesian information criterion (BIC). Machine learning algorithms are popular to classify quasars from the other objects these days \citep{Richards+2004, Jin+2019, Nakoneczny+2019, Schindler+2019}, with multiple strong points. (1) Machine learning can quickly judge each astronomical object \citep{Gupta+2014}. (2) unlike linear color-cuts defined arbitrarily by the human inspection of the color space, it can optimize a nonlinear boundary mathematically by minimizing the difference between the true label and the predicted label from the trained model \citep{Kojima+2020}. (3) It can consider the estimates from all the bands as its input to decide the boundary between quasars and other astronomical objects, while the traditional color selection can utilize a few broadbands only. Thus, DL can perform a fast selection of quasar candidates with maximal completeness. The BIC selection is additionally implemented to refine the selected candidates and can remove contaminating sources efficiently (e.g., \citealt{Shin+2020}). 

\indent This paper is structured as follows. We described the HSC-SSP data in Section~\ref{sec:Data}, and training data for DL and quasar/star model for BIC in Section~\ref{sec:Models}. Section~\ref{sec:Method} describes the DL and BIC selection of quasars and the selected candidates. In Section~\ref{sec:Result}, we show how the quasar binned LFs and parametric LF are derived based on the final candidates. In Section~\ref{sec:Discussion}, we discuss how our quasar selection and the quasar LF compare with previous studies and how the improvement over previous works was possible. Section~\ref{sec:Summary} summarizes the findings and the results of this study. Throughout this paper, the AB magnitude system is adopted for all filters \citep{Oke+1983}, after the Galactic extinction correction by adopting the dust map of the \citet{Schlegel+1998}. We assume $\Omega_{M} = 0.3$, $\Omega_{\Lambda} = 0.7$ and $H_{0} = 70$ km s$^{-1}$ Mpc$^{-1}$ of the $\Lambda$CDM cosmology, which has been supported by observations in the past decades (e.g., \citealp{Im+1997})

\bigskip
\section{HSC-SSP Deep-layer catalog} \label{sec:Data}
We used the catalog constructed from the Deep layer of the HSC-SSP in PDR2 with a survey area of 27 deg$^2$ and a 5-$\sigma$ image depth of $\sim$ 27 mag for a point source in $r$-band \citep{Aihara+2019}. The Deep layer consists of four fields \citep{Aihara+2018}: the XMM Large-Scale Structure Survey (XMM-LSS, \citealt{Pierre+2004}), Extended-COSMOS (E-COSMOS, \citealt{Scoville+2007}), the European Large-Area {\it ISO} Survey-North 1 (ELAIS-N1, \citealt{Rowan-Robinson+2004}), and the DEEP2-3 \citep{Cooper+2011, Newman+2013}. We used the data taken in five broadbands ($g, r, i, z, y$) and two narrow-bands ($NB816, NB921$). The 5-$\sigma$ image depths of the seven bands ($g, r, i, z, y, NB816, NB921$) are (27.3, 26.9, 26.7, 26.3, 25.3, 26.1, 25.9) mags for a point source detection, respectively \citep{Aihara+2019}. An effective survey area of this layer in PDR2 is about 15.5 $\mathrm{deg}^{2}$, calculated from a random source catalog provided in the HSC data archive system \citep{Coupon+2018}.

\begin{deluxetable}{lc} 
\centering
\caption{The conditions used for retrieving sources} \label{tab:flags}
\tablehead{\colhead{Flag} & \colhead{Value} } 
\startdata
inputcount$\_$value & $\geqq2$ \\
detect$\_$primary & True\\
Localbackground$\_$flag$\_$nogoodpixels & False \\
pixelflags$\_$edge & False \\
pixelflags$\_$saturatedcenter & False \\
pixelflags$\_$crcenter & False \\
pixelflags$\_$bad & False \\
\hline
\enddata
\end{deluxetable}

\begin{figure*} 
\centering 
\includegraphics[width=0.95\textwidth]{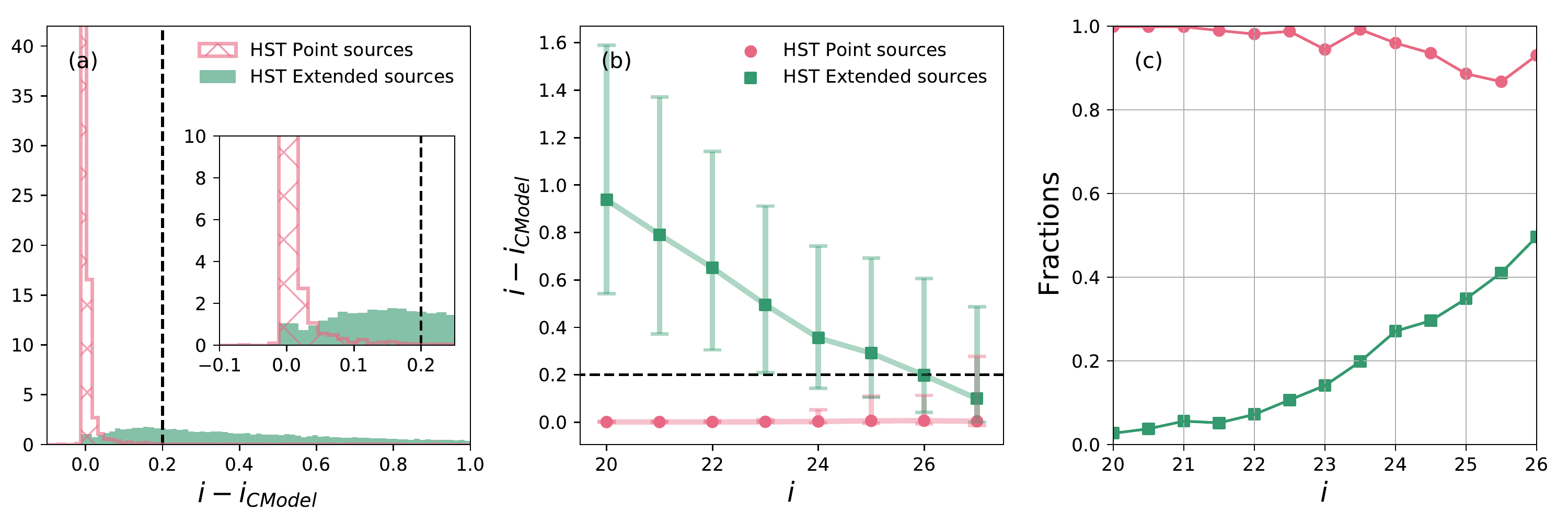}
\caption{(a) Normalized histograms of the extendedness values $(i - i_{\rm CModel})$ for the HST extended (green filled) and point sources (pink hatched). The black dashed line indicates $(i - i_{\rm CModel}) = 0.2$. $98.4\%$ of the HST point sources have $(i - i_{\rm CModel}) < 0.2$, whereas the $(i - i_{\rm CModel})$ values of the HST extended sources are almost evenly distributed. (b) The $(i - i_{\rm CModel})$ value distributions as a function of $i$-band magnitude for the HST extended (green squares and line) and point sources (pink circles and line). The median values of $(i - i_{\rm CModel})$ distributions are marked with squares or circles. The error bars indicate the 68$\%$ percentile of the distributions. The black dashed line indicates the extendedness value we applied to our quasar selection. (c) The fractions of the HST extended/point sources with $(i - i_{\rm CModel}) < 0.2$. The green squares indicate the point source contamination rate caused by galaxies, and the pink circles represent the point source completeness. Although the point source contamination reaches about $40\%$ at $i<25$ mag, the completeness is always $\gtrsim 90\%$.}
\label{fig:1_Pscut}
\end{figure*}

\indent We used the source catalog from the HSC-SSP PDR2 excluding objects whose photometry measurements are flagged to be affected by the cosmic ray, saturation, abnormal local background estimation, bad pixels, shallow depth, and the proximity to the survey edges. Also, we considered the objects that are primary and unique sources having no child in the survey. Table~\ref{tab:flags} lists the flags that we adopted to retrieve the catalog sources. The number of retrieved sources is about 3.5 million. For our analysis, we used the point-spread function (PSF) magnitudes (see Section~\ref{sec:pre-selection}).

\bigskip
\section{Training $\&$ Test Data$/$Models} \label{sec:Models}
\subsection{Training $\&$ Test Data} \label{sec:dataset}
To select reliable high-redshift quasar candidates using the DL technique, we should prepare a dataset including training data representing quasars at $z\sim5$ (‘qso’) and the other objects (‘nqso’), and data for testing the trained model. The ‘nqso’ class includes HSC-SSP sources satisfying our point source selection criterion. Although ‘nqso’ class might contain few real quasar samples, the probability of including real quasars is too low ($\sim 0.2 \%$, please refer to Section~\ref{sec:pre-selection}). This kind of empirical approach to constructing the ‘nqso’ sample has been used in previous works (e.g., \citealt{Timlin+2018}) when the properties of the ‘nqso’ population are poorly known. The ‘qso’ dataset is made up of quasar SED models (described in Section~\ref{sec:quasar_model}) because of a small number of spectroscopically confirmed quasars at $z\sim5$ compared to the other classes. The training dataset for each class is a randomly sampled subset, as described in Section~\ref{sec:training}.

\indent We find six spectroscopically identified quasars and three promising quasar candidates at $4.5 <$ z $< 5.5$ in the Deep layer \citep{McGreer+2013, Paris+2018, Shin+2020} with a matching radius of $1\farcs0$. The matched quasars are used for testing the performance of the trained model independently. 

\subsection{Quasar/Star SED model}
\subsubsection{Quasar} \label{sec:quasar_model}
We created model spectra of quasars at $z\sim5$ by creating a composite SED of \citet{Lusso+2015} and \citet{Selsing+2016} at the bluewards and redwards of 1450 \AA, respectively. Compared to the composite quasar spectrum of \citet{VandenBerk+2001}, their SEDs are more likely to be intrinsic ones, free from UV absorption and host galaxy contamination. 

\indent Then, we manipulated the equivalent width (EW) of Ly$\alpha$ and N \Romannum{5} $\lambda 1240$ and the continuum slope ($\alpha_{\lambda}$) of the model following their empirical distributions of high-redshift quasars; a log-normal distribution with $\log{\mathrm{EW}} = 1.524 \pm 0.391$ \citep{Banados+2016} and $\alpha_{\lambda} = -1.6 \pm 1.0$ \citep{Mazzucchelli+2017}. Concerning the IGM absorption at high redshifts, we used an updated version of the IGM attenuation model \citep{Inoue+2014}. Finally, we rescaled the EW of the C\Romannum{4} emission line by multiplying the rate of the EW change of Ly$\alpha$ and N \Romannum{5} $\lambda 1240$. The Baldwin effect was not considered in this study \citep{Baldwin+1977}, not to bias our sample to be those that follow the Baldwin effect. 

\indent The quasar SED model has four parameters: the redshift (z), EW, $\alpha_{\lambda}$, and $M_{\rm 1450}$. The redshift range of the model is 4.0 to 6.0 in steps of $\Delta$z $= 0.01$, and the $\alpha_{\lambda}$ range is -3.6 to 1.6 in steps of $\Delta\alpha_{\lambda} = 0.2$. The log EW (\AA) grid consists of twenty evenly-spaced values from 0.76 to 2.324, and the $M_{\rm 1450}$ range is $-28 < M_{1450} < -20$ in steps of $\Delta M_{1450} = 0.1$ mag. The EW and $\alpha_{\lambda}$ ranges of the constructed quasar models enable us to account for 95.4 percent of the entire quasar population, respectively, assuming that quasars follow the $\alpha_{\lambda}$ and EW distributions. As a result, there are 12 million model spectra in total.

\subsubsection{Star} \label{sec:stellar_model}
We adopted the stellar model spectra generated by the BT-Settl models that use the ‘BT2’ water vapor line list computed in \citet{Barber&Tennyson+2006} and the ‘Settl’ model accounting for dust formation and its gravitational settling \citep{Allard+2003} based on solar abundances of \citet{Asplund+2009}. We used a total of 14,342 spectra covering the parameter space of $T_{\text{eff}}$ of 400--70,000 K with 50--100 K step sizes, log($g$) of -0.5--6.0 with a 0.5 step size, [M/H] of -4--0.5 with step sizes of 0.2 $\sim$ 0.5, and [$\alpha$/M] of 0.0--0.6 with a 0.2 step size. Note that [$\alpha$/M] varies only for the model with $T_{\text{eff}} > 2600 {\text{K}}$, and we added a normalization factor (f$_{N}$) as a free parameter.

\bigskip
\section{Quasar selection} \label{sec:Method}
We used multiple methods sequentially to select quasar candidates at $z\sim5$: pre-selection, DL, BIC, and visual inspection. The pre-selection picks out the candidates satisfying our survey design and the minimal conditions required to be a quasar, significantly curtailing the number of the candidates. DL plays the role of judging an object's class using the colors of the pre-selected candidates. Similar to DL, BIC indicates a likelihood of an object being a star or a quasar based on the results from the SED fitting. For candidates passing this BIC selection, we check the quality of images used for measuring their fluxes.

\subsection{Pre-selection} \label{sec:pre-selection}

We first selected candidates in the following order. (1) magnitude and error cuts were set to $19<i<25$ mag and $\sigma_{i} < 0.2$ mag, respectively; (2) we carried out a point source selection using the $i$-band parameter; and (3) we set a $g-r$ color cut to select red objects that are consistent with being at $z>4.5$. 

\indent First, we limited the $i$-band magnitude range from 19 to 25 mags, avoiding saturation and allowing to search quasars fainter as much as $M_{1450} \sim -21.0$ mag. Since quasars at $z\sim5$ are red objects with large $r-i \gtrsim 1.2$, the magnitude limit also enabled us to obtain $r$-band photometry above its detection limit of $r \sim 27$ mag. In addition, we set the error cut of $\sigma_{i} < 0.2$ to select sources with reliable $i$-band detection, eliminating few more objects ($< 0.01 \%$) that happen to be in a shallow regions of the HSC-SSP images. 

\indent Next, we distinguished point sources from extended sources. We included this process since a quasar whose light is dominated by an AGN at its center would appear as a point source. An obvious disadvantage of this selection is that we miss AGNs where host galaxies are more dominant, especially for AGNs with $M_{\rm UV} \gtrsim -23$ mag \citep{Trebitsch+2020, Bowler+2021, KimYJ+2021}. Therefore, our survey is limited to the AGNs that have strong emission lines and outshine their host galaxies in UV.

\indent The point source selection can be done by comparing PSF magnitudes and CModel magnitudes. We called the difference between the $i$-band PSF and CModel magnitudes ($i-i_{\rm CModel}$) the extendedness parameter and adopted it to classify point sources. While the extendedness parameter value is close to 0 for point sources, the parameter value deviates significantly from 0 for extended sources due to a mismatch in the object extendedness and the point source modeling. The extendedness parameter has been frequently adopted to address an extendedness of an object (e.g., \citealp{Matsuoka+2018_2}).

\indent To the extendedness value for the point source selection, we used the $I$-band catalog of the \emph{Hubble Space Telescope} (HST) Advanced Camera for Surveys (ACS, \citealt{Leauthaud+2007}). \citet{Leauthaud+2007} classified point sources in the surface brightness (\text{MU\_MAX}) versus apparent magnitude (\text{MAG\_AUTO}) plane, with a cut of \text{MU\_MAX} $<21.5$. Their point source classification was nearly complete at $I<25$ mag. Therefore, we matched the HST point/extended sources from \citet{Leauthaud+2007} to the HSC $i$-band sources at $i<25$ mag since our purpose is to search quasars as faint as $i\sim 25$ mag.

\indent Figure~\ref{fig:1_Pscut}a displays the histograms of point and extended sources as a function of $(i - i_{\rm CModel})$. In Figure~\ref{fig:1_Pscut}b, we show the dependency of the extendedness value of the HST point and extended sources on the $i$-band magnitude. The error bars correspond to the 16th and 84th percentiles of the distributions, respectively. As the $i$-band magnitude becomes fainter, they overlap more. Figure~\ref{fig:1_Pscut}c shows the point-source completeness and contamination rate of the HST point/extended sources with $(i - i_{\rm CModel}) < 0.2$ as a function of $i$-band magnitude. The point-source completeness is $\sim 90\%$ at $i=25$ mag, and 98.4 \% at $i<25$ mag, while the contamination rate reaches about $40\%$ at $i\sim25$ mag.

\indent Even though the point source selection cut caused a high contamination rate of an extended source mimicking a point source, we focused on increasing the point source completeness. The following selections using SED shape can further weed out high-redshift objects that are dominated by galaxy light.

\indent In the third step, we eliminated the sources with the possible detections at the blueward wavelengths, which are clearly not high-redshift quasars with the IGM attenuation. We selected the sources that are not detected in $g$-band imposing $\mathrm{SNR}<3$, or had $g-r>0.987$, which is determined from the minimum $g-r$ color of our quasar model at $z=4.5$ (Section~\ref{sec:quasar_model}). After these steps, the number of pre-selected objects is 125,644 over 15.5 deg$^2$.


\subsection{Deep Learning}

\subsubsection{A brief introduction}
The first mathematical expression about a neural network was introduced by \citet{Mcculloch+1943}. The successful performance of a convolutional neural network in the ImageNet project proved the potential of the neural network \citep{imagenet_cvpr09}, and it spurred the application of DL to many other disciplines including astronomy. 

\indent A neural network consists of multiple layers: an input layer, an output layer, and hidden layers. The input layer receives an object's information, and the output layer returns the object's property we want to know. The hidden layers connect the information in the input layer to the output object's property. In this work, we implemented a supervised DL to predict a class for an object (output) based on its photometry (input). 

\indent Each layer has neurons which are the smallest data-processing units. Each neuron in each layer except for the input layer has a weight, $w$, corresponding to each feature of an input, $x$, and a bias, $b$. The neuron calculates the weighted sum of the features, adds the bias to the sum, and passes the sum to an activation function, $A$. The activation functions rescale the sum and determine whether the output value, $y$, for each neuron should be activated or not ($y=0$). The following equation shows how to calculate the output value for an $i$-th neuron in an $l$-th layer.

\begin{equation}
\begin{aligned}
    y_{i}^{l} = A\; (\; \sum_{j=1}^{n_{l-1}} [w_{i,j}^{l}x_{j}^{l-1}] + b_{i}^{l} \;), \\
\end{aligned}
\end{equation}
where $n_{l-1}$ is the number of neurons in the ($l-1$)th layer. The output of the $i$-th neuron in the $l$-th layer ($y_{i}^{l}$) becomes the input to the neurons in the next layer. If the next layer is an output layer, then the output becomes the probabilities. Weights and a bias for each neuron contribute to predicting the final outputs. Thus, the main purpose of the model training is to find appropriate weights and a bias for each neuron. The optimal model parameters ($w$ and $b$) can be obtained by minimizing the loss function, which considers the difference between the true class and the predicted class by the model. 

\begin{deluxetable*}{cccccc}[th] 
\caption{Hyperparameter search space} \label{tab:hyperparameter}
\tablehead{\colhead{Hyperparameter} & \colhead{} & \colhead{Search space} & \colhead{} } 
\startdata
The number of neurons in each hidden layer & & [10,20,30] & \\
Weight decay & & uniform distribution from 1e-5 to 1e-4 & \\
Learning rate & & uniform distribution from 1e-4 to 1e-2 & \\
Momentum & & unifrom distribution from 0.7 to 1 & \\
Batch size & & [32, 64, 128, 256] & \\
Epochs && $<20$ & \\
Initial values for $w$ && normal distribution & \\
\hline
\enddata
\end{deluxetable*}

\subsubsection{Hyperparameter optimization}
\indent Before optimizing the weights and bias of each neuron, we examined the best combination of the hyperparameters. The hyperparameters are the parameters affecting the entire training process. It includes parameters related to the architecture of the neural network (the number of hidden layers and the number of neurons in each hidden layer), and rules for model training ($A$, loss function, weight decay, optimizer, batch size, epochs, initial values for $w$). The hyperparameter combination can influence the converge time for finding the optimal model parameters and the model performance. For efficient model training, we have to find the best configuration of the hyperparameters. The hyperparameters are explained in detail:

\begin{itemize}
    \item The number of hidden layers
    \item The number of neurons in each hidden layer
    
    \item Activation function: Each neuron has an activation function to decide whether the inputs contribute to minimizing the loss function or not. The activation function also enables us to calculate the gradient of the loss function with respect to the weights and the bias of each neuron according to the Backpropagation algorithm \citep{Rumelhart+1986}. Among various activation functions, we adopted the Rectified Linear Unit (ReLU, \citealt{Nair+2010}), given as $f(x)$ == max$\{0,x\}$.
        
    \item Loss function: The difference between the true and the predicted properties for a given model. We used the cross-entropy loss, which has been widely used due to the discrete property of the output, to calculate a mean loss of the training data in a batch.
    
    \item Weight decay: The penalty for the large weights. The weight decay term was additionally applied to the loss function to avoid overfitting.
    
    \item Optimizer: An optimization algorithm to minimize the loss function. We used stochastic gradient descent (SGD). The SGD calculates partial derivatives of the loss function of given weights or biases and updates the two parameters iteratively toward finding a global minimum. The SGD uses the randomly selected subsets of the training data (i.e., batch sample) at each iteration. 
    \begin{itemize}
        \item Learning rate: The step size of model parameters to explore the partial derivatives of the loss function
        \item Momentum: The fraction of taking into account the previous update to calculate the current update for a parameter. Adopting the momentum, we can accelerate convergence to the minimum by giving more weights to previous directions compared to the current direction which may be biased toward a noise.
    \end{itemize}
    
    
    \item Batch size: The number of training subsets used to calculate a gradient descent
    \item Epochs: The number of passing all training data to train the model
    \item Initial values for $w$: To implement the Backpropagation algorithm and update $w$ and $b$, we should assign initial values for $w$. In this paper, we randomly selected the initial values from a normal distribution in which the mean and standard deviation are 0 and 1. 
    
\end{itemize}

We constructed a feed-forward four-layer neural network. To prevent the risk of over-fitting, we set 20 epochs as an upper limit of the time for evaluating the derivatives and updating the model parameters. Except for the fixed hyperparameters (e.g., the number of hidden layers, activation function, loss function), we determined an optimal hyperparameter combination among the hyperparameter search spaces specified in Table~\ref{tab:hyperparameter}. For this, we used the Bayesian model-based optimization algorithm of the {\tt \string RayTune} Python package \citep{Bergstra+2013, Liaw+2018}, and tried 100 hyperparameter configurations. The hyperparameter set with the lowest loss was chosen as the final hyperparameter set for testing pre-selected sources in Section~\ref{sec:pre-selection}. 

\subsubsection{Preprocessing the inputs}
As the inputs of the training process, we used six colors from the catalog: $g-r, r-i, i-NB816, NB816-z, z-NB921, NB921-y$. When an $i$-band-selected source is not detected or fainter than the imaging depth of the other bands, we adopted 5-$\sigma$ imaging depths as their magnitudes in the corresponding bands. Considering the discriminative feature in $g$-band owing to the IGM absorption, we assigned $g=30$ mag when the object is not detected or has a magnitude fainter than 30 mag in $g$-band.

\indent After refining the magnitudes, we calculated the colors of the dataset, standardized each color, and extracted six principal components using the {\tt \string scikit-learn} Python package \citep{Pedregosa+2011}. The standardization removes the mean of each color and scales its standard deviation, enabling the principal component analysis (PCA) to weigh each color equally. Note that we used all the six principal components derived from the PCA, although the PCA is frequently used for dimension reduction of input features.

\begin{figure}[t]
\centering 
\includegraphics[width=0.49\textwidth]{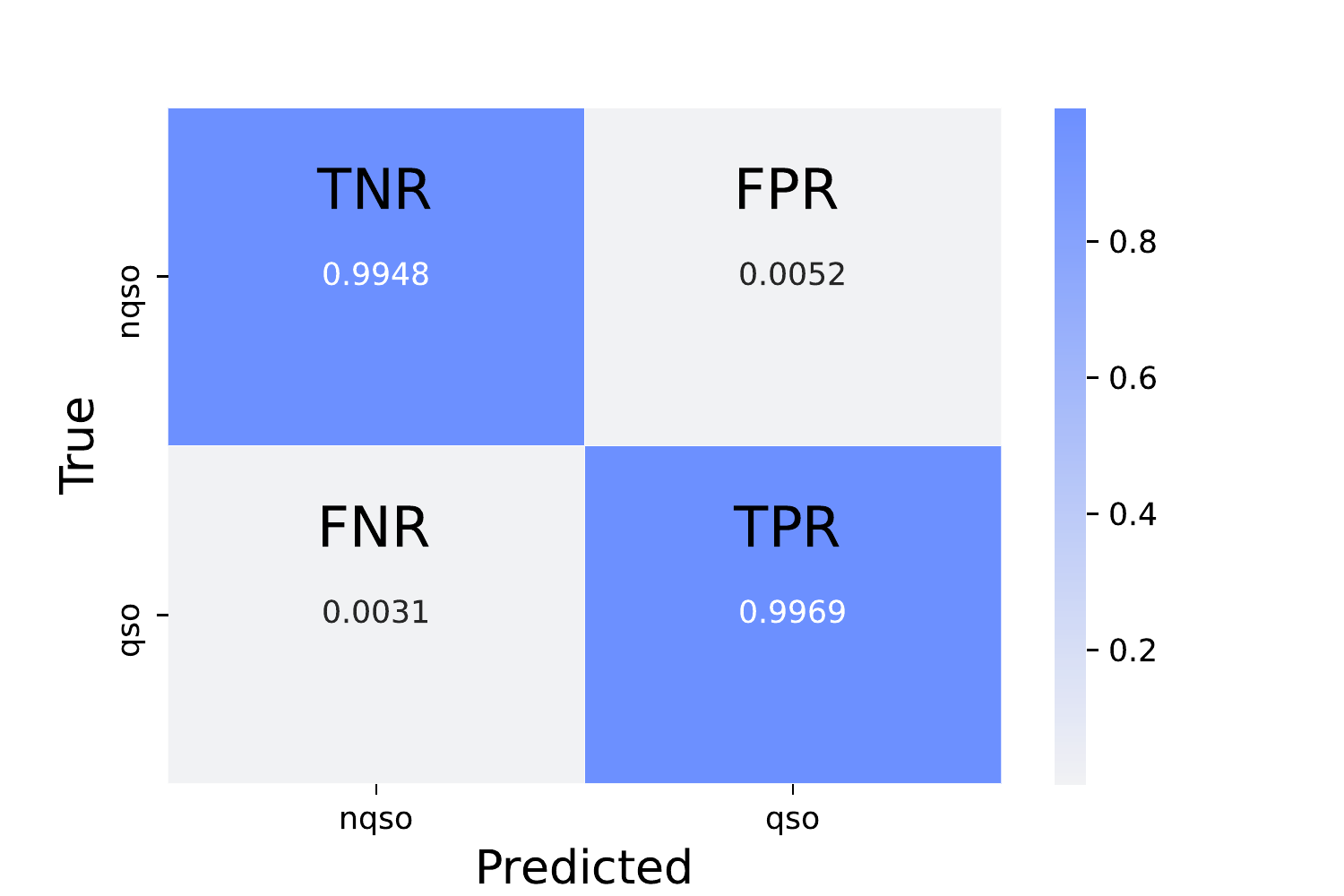}
\caption{Confusion matrix of our ensemble learning for the test dataset containing 20,000 nqso and 20,000 qso objects. The number in each section represents the fraction of the objects classified to a specific predicted label. The sum of the numbers in each row is 1.0.}
\label{fig:2_ConfMat}
\end{figure}

\subsubsection{Training} \label{sec:training}
Our training set contains two classes of objects -- the ‘nqso’ class and the ‘qso’ class. The ‘nqso’ class denotes objects that are not quasars and the nqso class set consists of about $3.34\times10^{5}$ $i$-band-selected point sources in the Deep layer of HSC-SSP. The ‘qso’ class denotes quasars, and the qso class set consists of millions of quasar model SEDs at $z=4.5-5.5$ (Section~\ref{sec:quasar_model}). The class imbalance problem is handled by randomly sampling the nqso and the qso classes when making a total sample \citep{Buda+2017}. In reality, the nqso contains real quasars in the survey area. However, given that the number of real quasars is expected to be small (a few tens), their contamination of the nqso sample is negligible. The number of sources in each class was fixed to 100,000, resulting in a total dataset size of 200,000.

\indent We set aside 20 $\%$ of the dataset as the test dataset. The remainder was split into five subsets -- four for the training dataset and one for the validation set. To minimize possible dependence on a given training dataset, we performed a 5-fold cross-validation by changing the subset used for the validation set. If DL classified an object as qso more than or equal to 3 times, we considered the object as a quasar candidate of the trained model. 

\indent Also, to make our selection more robust, we trained additional 99 neural network models by following the above procedure \citep{Durovcikova+2020}. From the 100 results of the 100 models, we classified an object as qso if the DL-selected candidates show qso label more than or equal to 80 times. Figure~\ref{fig:2_ConfMat} shows the confusion matrix for our DL selection. The probability of an actual nqso object to be predicted as qso class (False Positive Rate, FPR) is extremely low ($\sim 0.5 \%$). The probability is also as low as $\sim 0.3 \%$ for the trained model classifying an actual qso object as nqso class (False Negative Rate, FNR). The low FNR can assure us high completeness of the quasar survey. In Section~\ref{sec:comp_cc_vs_DL}, we compare previous quasar selections and the DL selection. The number of quasar candidates from the ensemble learning is 1,599. 


\subsection{SED fitting for BIC selection}

Although FPR is very low, misclassified nqso objects could occupy a large portion of the quasar candidates because the absolute number of nqso objects is larger than that of qso objects in a real world ($\sim 10,000:1$). To remove misclassified nqso objects from quasar candidates, we performed the SED fitting and an additional BIC selection, as in \citet{Shin+2020}. We briefly summarize the procedure as follows.

\indent First, we fitted the SED of the DL-selected quasar candidates with both quasar and star models. To allow margins for errors, we chose a redshift range spanning 4.0 to 6.0 for the quasar model fit, a bit broader than the redshift range of the model used for DL. Throughout the SED-fitting process, we adopted the chi-square calculation presented in \citet{Sawicki+2012}, which deals with the upper limits of observation data. The fitting results provided the chi-square values and the best-fit parameters of the quasar and star models. We excluded the candidates with the best-fit quasar model of $\chi^{2}_{qso} >$ 30.

\indent Then, we calculated BIC, a criterion used for model selection considering a likelihood and the number of free parameters in a model, $k$. In general, a fitting result becomes better as $k$ increases. Giving a penalty to a model with many parameters, the difference between the BIC values of different models ($\Delta {\rm BIC}$) can determine a preferred model. It is defined as 

    \begin{equation}
    \Delta\text{BIC} = (\chi^2_{\text{star}} - \chi^2_{\text{qso}}) + (k_{\text{star}} - k_{\text{qso}}) \times \ln n,
    \end{equation}

\noindent where $n$ is the number of data points of an object, and $\chi^{2}$ is the chi-square value of a best-fit model. The ‘star’ and ‘qso’ subscripts mean the best-fit model for star and quasar. If $\Delta\text{BIC}$ of a DL-selected candidate is greater than 10, we regarded the candidate as the BIC-selected candidate \citep{Liddle+2007}.

\indent After the BIC selection, the number of quasar candidates becomes 78. Note that one of three promising candidates reported in \citet{Shin+2020} and a known quasar at $z=4.564$ were excluded in this process. We discuss this issue in~\ref{sec:dis2}.

\subsection{Visual inspection} \label{sec:visins}
We visually inspected images of the 78 candidates and excluded 25 of them due to spurious photometry results caused by bright neighbors, background variations, optical ghosts, satellite tracks, or scattered lights. These features caused the local background overestimates, resulting in flux underestimates in $g$-band that mimicked the redshifted Lyman break. 

\indent Then, we examined if the remaining 53 candidates were previously reported by querying NASA/IPAC Extragalactic Database (NED) using the {\tt \string astroquery}. We recovered 5/6 confirmed quasars \citep{McGreer+2013, Shin+2020} and two promising candidates in \citet{Shin+2020}. We also recovered a candidate with a probability to be a quasar P$_{\rm qso} = 1$ in \citet{McGreer+2018}, CFHTLS J021800.49-044718.5, and another possible AGN at $z=4.549$ in \citet{Chaves-Montero+2017}, ALH3L490. In addition, our candidates include a spectroscopically confirmed galaxy at z $\sim$ 5 (\citealt{Ono+2018}, refer to Section~\ref{sec:dis1}).   

\subsection{Final candidates} \label{sec:FinalCan}

\begin{figure}[t!]
\centering 
\includegraphics[width=0.49\textwidth]{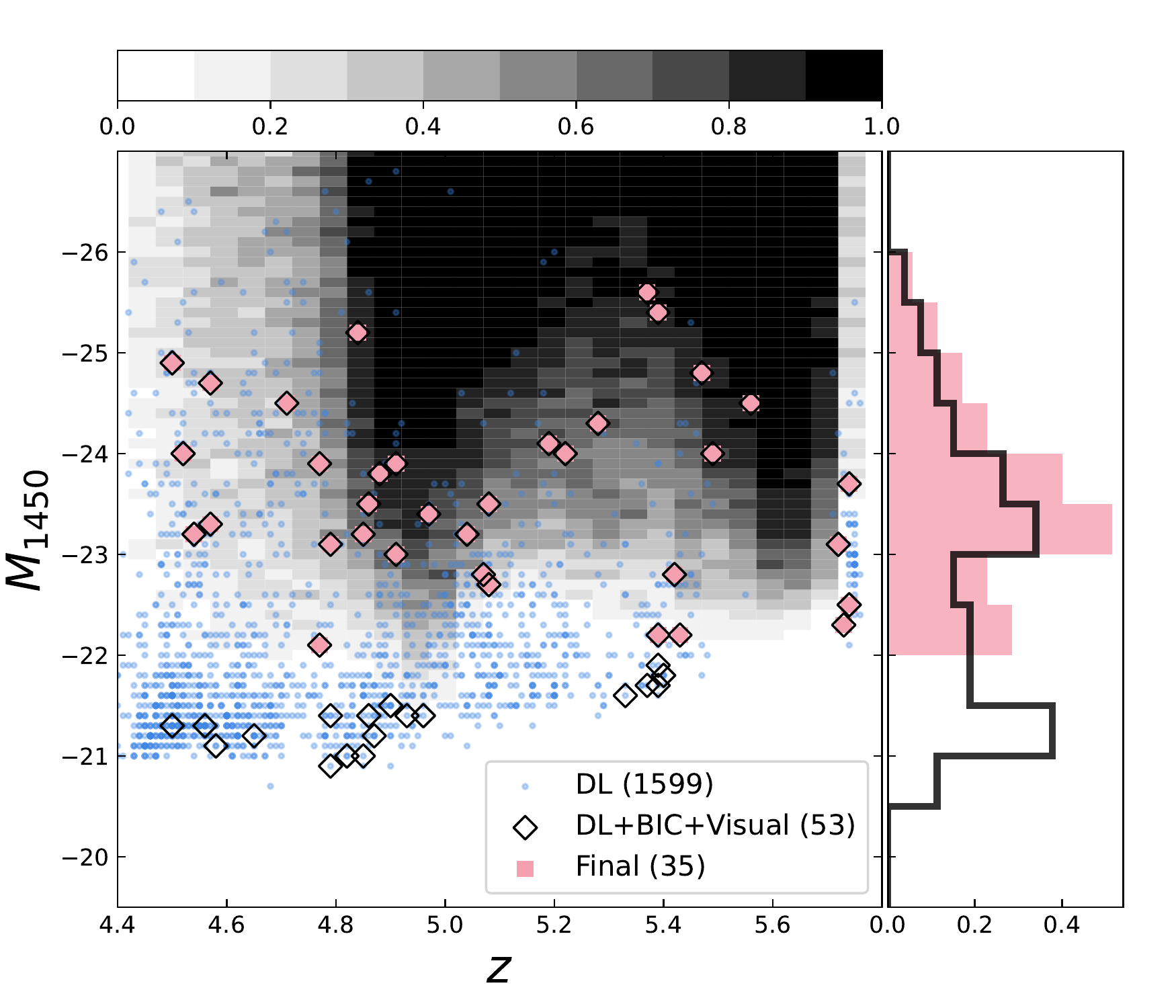}
\caption{The $z_{\rm phot}$ and $M_{1450}$ distributions of candidates selected by each process, after the initial selection. The candidates satisfying DL criteria (the blue circles), a combination of DL, BIC, and visual inspection (the black diamonds), and all the selection criteria (the pink squares) are marked. The total completeness function as a function of $z$ and $M_{1450}$ is expressed as a two-dimensional histogram. This function has a value from 0 to 1, indicating the fraction of quasar SED models satisfying pre-, DL and BIC selections in a bin with $\Delta z=0.05$ and $\Delta M_{1450}=0.1$. The sided panel shows the normalized $M_{1450}$ histogram.}
\label{fig:3_CompwCans}
\end{figure}

To assess whether a visually inspected candidate was realistic or not, we checked the probability of finding each candidate in our survey. First, we made a completeness function shown in Figure~\ref{fig:3_CompwCans}, $F(z, M_{\rm 1450})$, of our survey. This function is the fraction of the quasars satisfying the pre-, DL, and $\Delta$BIC selections among our simulated quasars within given bin sizes of $\Delta M_{1450}$ = 0.1 mag and $\Delta$z = 0.05. One hundred quasar models are in each bin. Note that we gave additional errors to the model magnitudes. The errors were randomly sampled from a normal distribution determined from the model magnitude and the imaging depths.

\begin{figure*} 
\centering 
\includegraphics[width=0.95\textwidth]{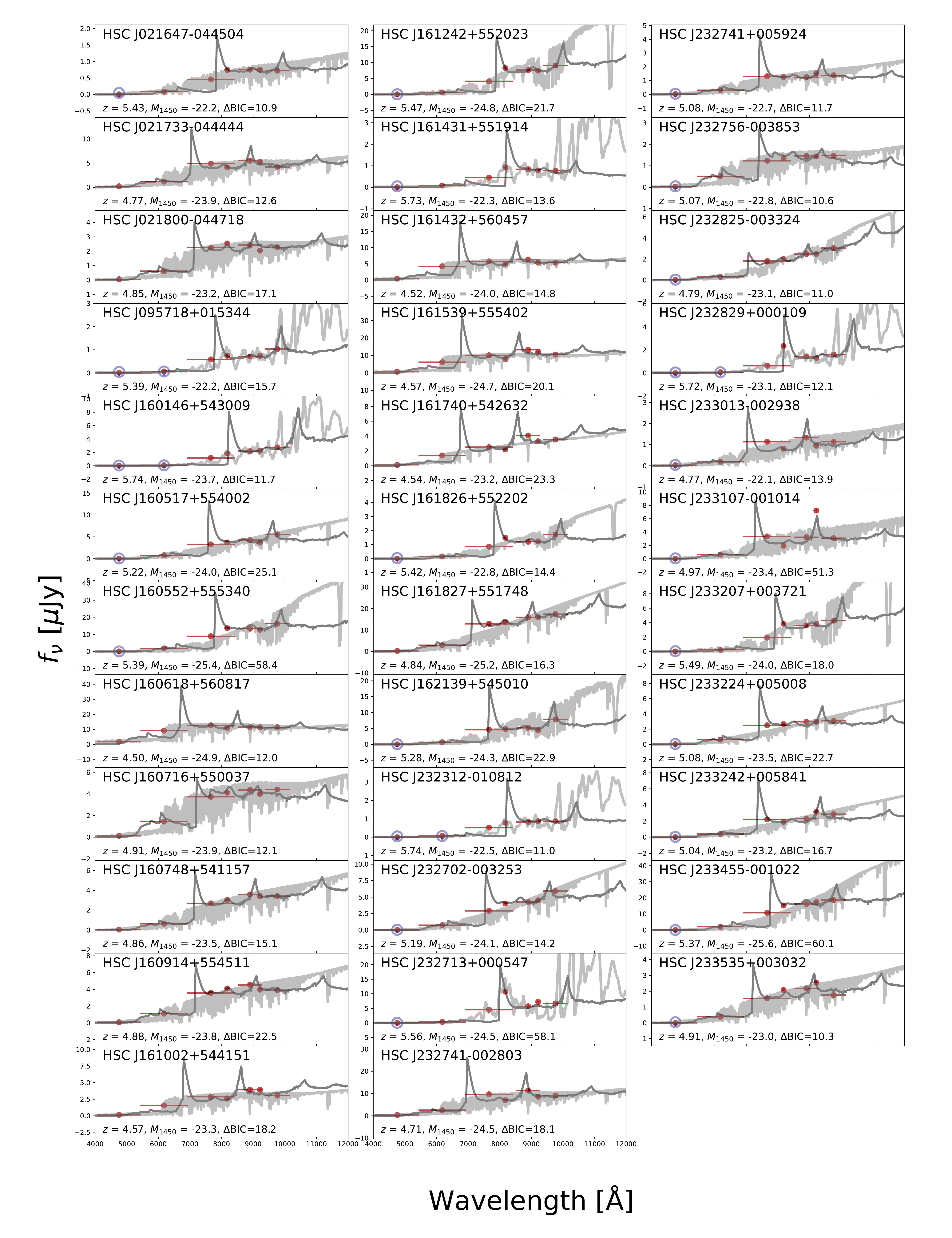}
\caption{The SED-fitting results of 35 final candidates. The best-fit quasar and star SED models are plotted with dark gray and light gray lines. The red points show the fluxes measured in the HSC-SSP filters. The blue upper limit means the non-detection case or a flux value being fainter than the 5-$\sigma$ limiting magnitude for a point source. Three of the best-fit parameters of quasar models ($z, M_{1450}$, and $\Delta\text{BIC}$) are provided in the lower part of each panel.}
\label{fig:4_SED}
\end{figure*}

\indent Figure \ref{fig:3_CompwCans} shows that the 53 quasar candidates selected with the DL+BIC+Visual inspection have a bimodal distribution, with a peak in their numbers at $M_{1450} \sim -21$ mag and another peak at $M_{1450} < -22.0$ mag. Considering that the quasar selection completeness is very low at $M_{1450} > -22$ mag, the large fraction of faint quasar candidates at $M_{1450} > -22$ mag are likely to be contaminated by z $\sim$ 5 galaxies that are known to be the dominant population at those magnitudes. To remove galaxy interlopers, we considered the candidates with $M_{1450} < -22.0$ mag as our final quasar candidates. The number of the final candidates is 35 with $\langle z_{\rm phot} \rangle \sim 5.0$, excluding a spectroscopically confirmed galaxy at z $\sim$ 5 \citep{Ono+2018} mentioned in Section~\ref{sec:visins}. The HSC-SSP photometry of 35 final candidates is listed in Table~\ref{tab:photometry} and their SED-fitting results are plotted in Figure~\ref{fig:4_SED}.

\begin{deluxetable*}{cccccccccccccccc} 
\caption{35 final quasar candidates at $z\sim5$} \label{tab:photometry}
\tablewidth{550pt}
\tabletypesize{\scriptsize}
\tablehead{\colhead{R.A.(J2000)} & \colhead{Decl.(J2000)} & \colhead{$g$} & \colhead{$r$} & \colhead{$i$} & \colhead{$NB816$} & \colhead{$z$}
& \colhead{$NB921$} & \colhead{$y$} & \colhead{$M_{1450}$} & \colhead{$z_{\rm phot}$} & \colhead{$z_{\rm spec}$} & \colhead{Notes}}
\startdata
02:16:47.45 & -04:45:04.1 & $>$ 27.30 & 26.66 & 24.75 & 24.22 & 24.21 & 24.21 & 24.25 & -22.2 & 5.43 & &  \\ 
02:17:33.44 & -04:44:44.3 & 25.71 & 23.73 & 22.18 & 22.36 & 22.05 & 22.10 & 22.34 & -23.9 & 4.77 & &  \\ 
02:18:00.51 & -04:47:18.6 & 26.93 & 24.43 & 23.02 & 22.89 & 22.94 & 23.13 & 23.01 & -23.2 & 4.85 & & \citealt{McGreer+2018} \\ 
09:57:18.73 & +01:53:44.3 & $>$ 27.30 & 26.87 & 24.48 & 24.22 & 24.27 & 24.22 & 23.87 & -22.2 & 5.39 & &  \\ 
16:01:46.59 & +54:30:09.9 & $>$ 27.30 & 26.69 & 23.72 & 23.23 & 23.04 & 23.03 & 22.80 & -23.7 & 5.74 & &  \\ 
16:05:17.79 & +55:40:02.0 & $>$ 27.30 & 24.18 & 22.62 & 22.48 & 22.33 & 22.48 & 22.04 & -24.0 & 5.22 & 5.211 & \citealt{Shin+2020} \\ 
16:05:52.06 & +55:53:40.6 & $>$ 27.30 & 23.27 & 21.52 & 21.06 & 21.08 & 21.14 & 20.87 & -25.4 & 5.39 & 5.409 & \citealt{Shin+2020} \\ 
16:06:18.73 & +56:08:17.3 & 23.29 & 21.50 & 21.14 & 21.29 & 21.24 & 21.26 & 21.26 & -24.9 & 4.5 & &  \\ 
16:07:16.10 & +55:00:37.5 & 26.19 & 23.50 & 22.47 & 22.36 & 22.30 & 22.39 & 22.29 & -23.9 & 4.91 & &  \\ 
16:07:48.14 & +54:11:57.4 & 27.07 & 24.40 & 22.83 & 22.70 & 22.52 & 22.57 & 22.57 & -23.5 & 4.86 & & \citealt{Shin+2020} \\ 
16:09:14.68 & +55:45:11.7 & 26.57 & 23.79 & 22.52 & 22.37 & 22.26 & 22.41 & 22.42 & -23.8 & 4.88 & 4.814 & \citealt{Shin+2020} \\ 
16:10:02.01 & +54:41:51.9 & 26.10 & 23.42 & 22.76 & 22.86 & 22.41 & 22.42 & 22.69 & -23.3 & 4.57 & & \citealt{Chaves-Montero+2017} \\ 
16:12:42.10 & +55:20:23.6 & $>$ 27.30 & 24.46 & 22.36 & 21.61 & 21.69 & 21.72 & 21.51 & -24.8 & 5.47 & &  \\ 
16:14:31.25 & +55:19:14.4 & $>$ 27.30 & 26.66 & 24.77 & 23.99 & 24.09 & 24.17 & 24.18 & -22.3 & 5.73 & &  \\ 
16:14:32.19 & +56:04:57.5 & 24.76 & 22.33 & 22.00 & 22.14 & 21.89 & 22.07 & 22.08 & -24.0 & 4.52 & &  \\ 
16:15:39.23 & +55:54:02.2 & 24.25 & 21.92 & 21.39 & 21.67 & 21.10 & 21.20 & 21.34 & -24.7 & 4.57 & &  \\ 
16:17:40.94 & +54:26:32.4 & 26.19 & 23.54 & 22.90 & 23.04 & 22.37 & 22.60 & 22.53 & -23.2 & 4.54 & &  \\ 
16:18:26.35 & +55:22:02.6 & $>$ 27.30 & 25.95 & 24.08 & 23.46 & 23.70 & 23.68 & 23.31 & -22.8 & 5.42 & &  \\ 
16:18:27.29 & +55:17:48.5 & 25.34 & 22.73 & 21.12 & 21.07 & 20.90 & 20.89 & 20.80 & -25.2 & 4.84 & & \citealt{Shin+2020} \\ 
16:21:39.62 & +54:50:10.9 & $>$ 27.30 & 24.37 & 22.24 & 22.19 & 22.11 & 22.29 & 21.66 & -24.3 & 5.28 & &  \\ 
23:23:12.47 & -01:08:12.6 & $>$ 27.30 & 26.77 & 24.60 & 24.16 & 24.10 & 24.05 & 24.05 & -22.5 & 5.74 & &  \\ 
23:27:02.65 & -00:32:53.3 & $>$ 27.30 & 24.22 & 22.74 & 22.39 & 22.33 & 22.26 & 21.97 & -24.1 & 5.19 & &  \\ 
23:27:13.22 & +00:05:47.9 & $>$ 27.30 & 25.14 & 22.26 & 21.32 & 22.01 & 21.75 & 21.84 & -24.5 & 5.56 & &  \\ 
23:27:41.36 & -00:28:03.9 & 25.06 & 22.90 & 21.44 & 21.79 & 21.27 & 21.57 & 21.50 & -24.5 & 4.71 & 4.75 & \citealt{McGreer+2013} \\ 
23:27:41.94 & +00:59:24.4 & $>$ 27.30 & 25.17 & 23.60 & 23.65 & 23.66 & 23.44 & 23.54 & -22.7 & 5.08 & &  \\ 
23:27:56.35 & -00:38:53.1 & $>$ 27.30 & 24.62 & 23.67 & 23.58 & 23.49 & 23.51 & 23.48 & -22.8 & 5.07 & &  \\ 
23:28:25.33 & -00:33:24.4 & $>$ 27.30 & 25.18 & 23.26 & 23.15 & 22.91 & 22.91 & 22.69 & -23.1 & 4.79 & &  \\ 
23:28:29.21 & +00:01:09.8 & $>$ 27.30 & 26.87 & 24.41 & 22.98 & 23.51 & 23.62 & 23.39 & -23.1 & 5.72 & &  \\ 
23:30:13.23 & -00:29:38.7 & $>$ 27.30 & 25.72 & 23.76 & 24.13 & 23.59 & 23.96 & 23.76 & -22.1 & 4.77 & &  \\ 
23:31:07.00 & -00:10:14.5 & $>$ 27.30 & 24.44 & 22.59 & 23.18 & 22.63 & 21.75 & 22.68 & -23.4 & 4.97 & &  \\ 
23:32:07.75 & +00:37:21.8 & $>$ 27.30 & 25.48 & 23.19 & 22.44 & 22.51 & 22.44 & 22.32 & -24.0 & 5.49 & &  \\ 
23:32:24.60 & +00:50:08.9 & $>$ 27.30 & 24.41 & 22.90 & 22.84 & 22.72 & 22.72 & 22.67 & -23.5 & 5.08 & &  \\ 
23:32:42.24 & +00:58:41.0 & $>$ 27.30 & 24.89 & 23.04 & 23.21 & 22.99 & 22.65 & 22.75 & -23.2 & 5.04 & &  \\ 
23:34:55.06 & -00:10:22.2 & 27.03 & 23.13 & 21.33 & 20.94 & 20.88 & 20.80 & 20.73 & -25.6 & 5.37 & 5.11 & \citealt{McGreer+2013} \\ 
23:35:35.08 & +00:30:32.3 & $>$ 27.30 & 24.93 & 23.42 & 23.10 & 23.05 & 22.89 & 23.29 & -23.0 & 4.91 & &  \\ 
\enddata
\tablecomments{The magnitude errors are mostly less than 0.03 mag. The 5$\sigma$ detection limit is given with a sign of inequality if a candidate is not detected. For spectroscopically confirmed quasars, their spectroscopic redshifts ($z_{\rm spec}$) are provided.}
\end{deluxetable*}

\section{Quasar Luminosity Function at $z\sim5$} \label{sec:Result}
To construct the quasar LF at $z\sim5.0$, we assumed one redshift bin ranging $= 4.4 - 5.8$ and split the 49 visually inspected candidates with $-26 < M_{\rm 1450} < -21$ into six magnitude bins with $\Delta M_{\rm 1450} = 1.0$ or $0.5$ mag. It is worth noting that the binned LFs were calculated with 49 of the 53 visually-inspected candidates, only the final candidates of 35 were used for deriving a parametric LF due to the possible contamination from the high-redshift galaxies in the binned LFs with $M_{1450} > -22.0$ mag.

\indent To describe the bright end of the quasar LF, we used 96 bright quasars at $z\sim5$ from \citet{Yang+2016}. We redistributed the bright quasar sample to four $M_{1450}$ bins covering -28.5 to -27.0 mag, considering the differences between the adopted cosmological parameters in our and their works. In the same manner, we additionally secured quasars with moderate luminosity ($M_{1450} = -27.0$ to $-23.0$) from \citetalias{KimYJ+2020} to better determine the quasar LF.

\indent We calculated the effective survey volume using the updated 1/$V_{a}$ method \citep{Page&Carrera+2000}. In the original version of 1/$V_{a}$ method, the volume available to find the quasar at a given redshift range does not consider a dependency of the maximum detectable redshift on a given luminosity, while the updated 1/$V_{a}$ method does. Thus, the updated 1/$V_{a}$ method can estimate the survey volume accurately, especially for a faint magnitude bin near the detection limit of the survey. It is defined as, 

\begin{equation} \label{equ:V}
V = \frac{1}{\Delta M_{\rm1450}} \int_{\Delta M_{\rm1450}} \int_{z_{\rm min}}^{z_{\rm max}(M_{\rm 1450})} \! F(z, M_{\rm 1450})\, \frac{\mathrm{d}V}{\mathrm{d}z}\,\mathrm{d}z\,\mathrm{d} M_{\rm1450}\,,\\ \\
\end{equation}
where $z_{\rm min}$ is the lowest redshift of the redshift bin, and $z_{\rm max}(M_{\rm 1450})$ is the maximum redshift to discover quasars within a given magnitude bin. ${\mathrm{d}V}/ {\mathrm{d}z}$ is the cosmological volume element. $F(z, M_{\rm 1450})$ is the survey completeness defined in Section~\ref{sec:FinalCan}.

\indent The number density and its error corresponding to each magnitude bin were calculated using the following equations,
\begin{equation}
\Phi = \frac{N}{V\; \Delta M_{1450}}, \qquad \delta \Phi = \frac{\Phi}{\sqrt{N}}\\,
\end{equation}
where ‘$N$’ is the number of quasars or quasar candidates in a magnitude bin, and ‘$V$’ is the effective volume introduced in Equation~\ref{equ:V}. The uncertainty of $\Phi$ was estimated by the Poisson noise of $N$.

\begin{figure}
\centering 
\includegraphics[width=0.49\textwidth]{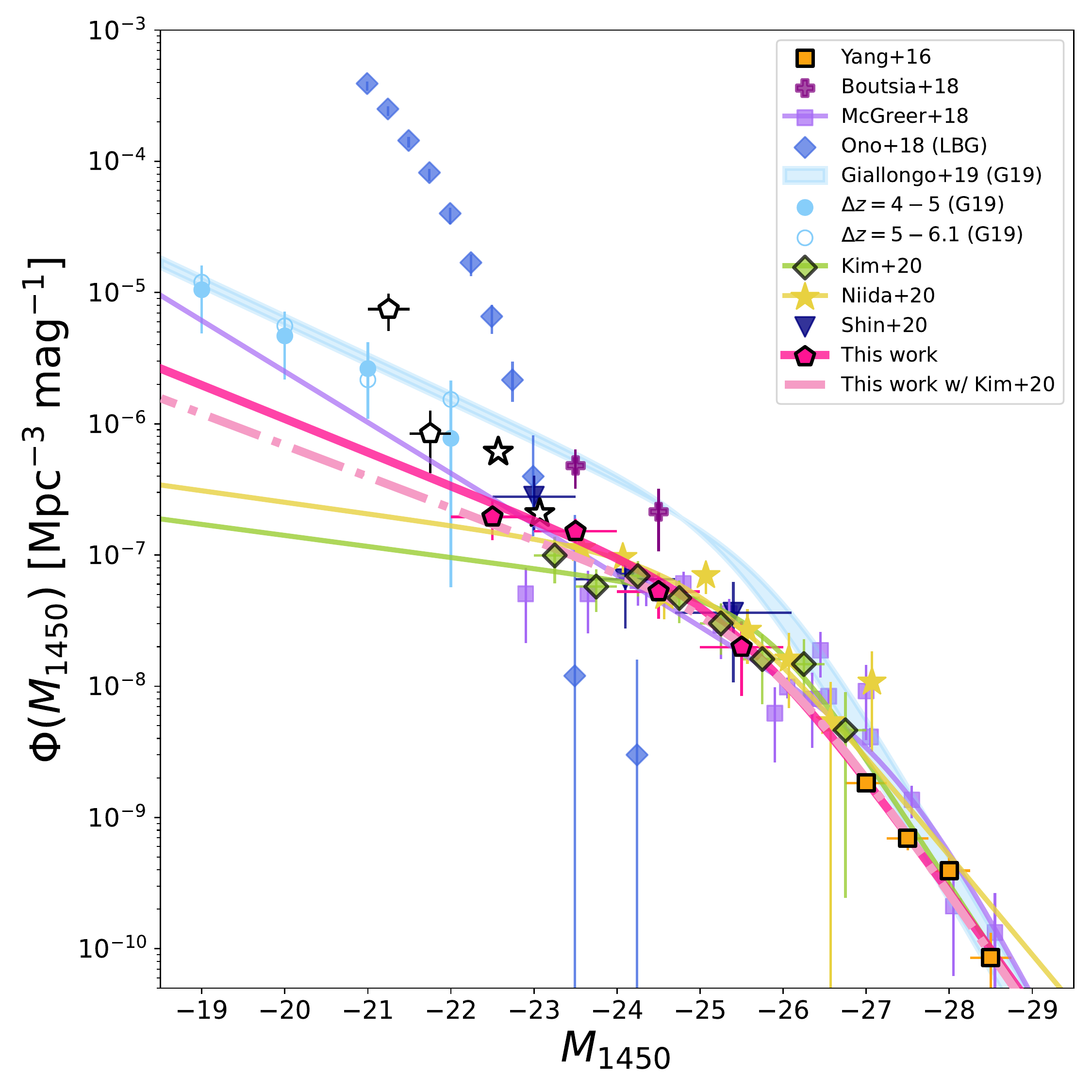}
\caption{Quasar LFs at $z\sim5$. The binned LF from our work is marked with hot pink pentagons. The re-binned LF of quasars in \citet{Yang+2016} and \citetalias{KimYJ+2020} are shown in the orange squares and the green diamonds, respectively. The pink solid (dash-dotted) line represents our parametric LF without (with) \citetalias{KimYJ+2020}. These binned and parametric LFs from our work are similar to those of the recently reported LFs (\citetalias{KimYJ+2020}, \citetalias{Niida+2020}, \citealt{Shen+2020}) at $M_{1450} < -23.5$. The empty pentagon and star points indicate the bins not used for obtaining a parametric quasar LF in this work and \citetalias{Niida+2020}, respectively, considering potentially high galaxy contamination. Contrary to several works, binned LFs of \citet{Boutsia+2018} and \citet{Giallongo+2019} show a high number density at the faint end.}
\label{fig:5_QLF}
\end{figure}

\indent We calculated the binned LFs based on the samples and survey completeness maps of \citet{Yang+2016}, \citetalias{KimYJ+2020}, and our survey. Except for ours, we re-scaled the binned LFs at z $\sim5.05$ to z $\sim5.0$ using the relation about the redshift evolution of number density at the break magnitude, $\Phi^{*}(\rm z) = \Phi^{*}(\rm z=6) \times 10^{k(z-6)}$ with k=-0.47 \citep{Fan+2001b}. Table~\ref{tab:binnedLF_Deep6} provides the binned LFs for quasars in \citet{Yang+2016} and \citetalias{KimYJ+2020}, and the 49 quasar candidates in our survey. Our binned LFs at $M_{1450} > -22.0$ mag increase dramatically as shown in Figure~\ref{fig:5_QLF}, implying possible galaxy contamination.

\indent To obtain the parametric quasar LF, we introduced a double power-law function of which the form is expressed as, 

\begin{equation}
\begin{aligned}
    & \Phi_{\rm model}(M_{\rm 1450})\  = \\
    & \frac{\Phi^{*}(\rm z=5.0)}{10^{0.4(\alpha+1)(M_{\rm 1450}-M^{*}_{\rm 1450})}+10^{0.4(\beta+1)(M_{\rm 1450}-M^{*}_{\rm 1450})}}.
\end{aligned}
\end{equation}

\indent Using the log-likelihood function defined as,

\begin{multline} \label{equ:S}
S = \sum_{\mathrm{ID}=1}^{n} \bigg( -2 \sum \ln [\Phi_{\rm model}(M_{1450})F_{\rm ID}(z, M_{\rm 1450})] \\
 + 2 \int \int \Phi_{\rm model}(M_{1450})F_{\rm ID}(z,M_{1450})\frac{\mathrm{d}V}{\mathrm{d}z}\,\mathrm{d}z\,\mathrm{d} M_{1450} \bigg), 
\end{multline}

\noindent where ID is a survey id corresponding to ours, \citet{Yang+2016}, or \citetalias{KimYJ+2020}, and $F_{\mathrm{ID}}$ is a completeness function of a survey with the ‘ID’. Giving a uniform prior to each parameter, we sampled the posterior distributions of the parameters by using the {\tt\string emcee} \citep{emcee+2013} python package to implement Markov Chain Monte Carlo (MCMC). The best-fit parameters and their uncertainties were determined from the 50th percentiles and 68$\%$ credible intervals of MCMC samples, respectively. The parametric LF was calculated using our 35 final quasar candidates, the bright quasars from \citet{Yang+2016}, and with or without moderate luminosity quasars from \citetalias{KimYJ+2020}.

\indent Table~\ref{tab:fittedLF_Deep6} summarizes the best-fit parameters of the parametric quasar LFs for whether the moderate luminosity quasars in \citetalias{KimYJ+2020} are included or not. Also, the fitted quasar LF model and binned LFs in this work are shown in Figure~\ref{fig:5_QLF}. Note that all the quasar LFs from the literature in the Figure are scaled to z=5.0.


%

\begin{deluxetable}{ccccc}[t] 
\caption{Quasar binned LF} \label{tab:binnedLF_Deep6}
\tablehead{\colhead{$M_{1450}$} & \colhead{$\Delta M_{1450}$} & \colhead{$\log{\Phi}$\tablenotemark{\textdagger}} & \colhead{$\delta\Phi$\tablenotemark{\textdagger}} & \colhead{$N$}} 
\startdata
\hline
\multicolumn{5}{c}{Re-binned LFs using quasars in \citet{Yang+2016}} \\
\hline
-28.5 & 0.5 & -10.093  & 0.047 & 3 \\
-28.0 & 0.5 & -9.429 & 0.099 & 14 \\
-27.5 & 0.5 & -9.182 & 0.131 & 25 \\
-27.0 & 0.5 & -8.761 & 0.236 & 54 \\
\hline
\multicolumn{5}{c}{Binned LFs using quasars in \citetalias{KimYJ+2020}} \\
\hline
-26.75 & 0.5 & -8.359 & 4.372 & 1  \\
-26.25 & 0.5 & -7.854 & 8.074 & 3  \\
-25.75 & 0.5 & -7.817 & 8.801 & 3  \\
-25.25 & 0.5 & -7.545 & 12.762 & 5  \\
-24.75 & 0.5 & -7.351 & 16.841 & 7  \\
-24.25 & 0.5 & -7.184 & 20.709 & 10 \\
-23.75 & 0.5 & -7.264 & 20.564 & 7  \\
-23.25 & 0.5 & -7.026 & 38.408 & 6  \\
\hline
\multicolumn{5}{c}{This work} \\
\hline
-25.50 & 1.0 & -7.702 & 11.463 & 3 \\
-24.50 & 1.0 & -7.279 & 19.889 & 7 \\
-23.50 & 1.0 & -6.820 & 37.830 & 16 \\
-22.50 & 1.0 & -6.710 & 64.999 & 9 \\
-21.75\tablenotemark{\textdaggerdbl} & 0.5 & -6.074 & 421.705 & 4 \\
-21.25\tablenotemark{\textdaggerdbl} & 0.5 & -5.127 & 2362.587 & 10
\enddata
\tablenotetext{$\textdagger$}{$\Phi$ is in units of Mpc$^{-3}$ mag$^{-1}$ and $\delta\Phi$ is in units of 10$^{-9}$ Mpc$^{-3}$ mag$^{-1}$.}
\tablenotetext{$\textdaggerdbl$}{These bins are highly contaminated by high-redshift galaxies (see Section~\ref{sec:FinalCan})}
\end{deluxetable}

\begin{deluxetable}{ccccccc}[t] 
\caption{Best-fit parameters of $\Phi_{\mathrm {model}}$} \label{tab:fittedLF_Deep6} 
\tablehead{\colhead{} & \colhead{${\log{\Phi^{*}}}$\tablenotemark{\textdagger}} & \colhead{$M^{*}_{\rm 1450}$} & \colhead{$\alpha$} &\colhead{$\beta$}} 
\startdata
w/o \citetalias{KimYJ+2020} & 
$-7.38^{+0.42}_{-0.58}$ &
$-25.55^{+0.75}_{-0.85}$ &
$-1.64^{+0.36}_{-0.30}$ &
$-3.22^{+0.24}_{-0.30}$ &\\
w/ \citetalias{KimYJ+2020} & 
$-7.56^{+0.26}_{-0.30}$ & 
$-25.83^{+0.47}_{-0.45}$ &
$-1.60^{+0.21}_{-0.19}$ &
$-3.32^{+0.22}_{-0.24}$ \\
\enddata
\tablenotetext{$\textdagger$}{$\Phi$ is in units of Mpc$^{-3}$ mag$^{-1}$}
\end{deluxetable}

\bigskip
\section{Discussion} \label{sec:Discussion}

\subsection{Comparison with previous LFs and implications on IGM ionization}
\indent As we explained earlier, we consider our quasar UV LF to be reliable down to $M_{1450} = -22.0$ mag, which goes about 1 mag deeper than the recent LFs (\citetalias{Niida+2020}, \citetalias{KimYJ+2020}). Here, we compare our LF with previous LFs in several aspects.

\indent We note that our two best-fit faint-ends ($\alpha \sim -1.64^{+0.36}_{-0.30}$ and $\sim -1.60^{+0.21}_{-0.19}$) are consistent with those of \citetalias{Niida+2020} ($\alpha \sim -2.0^{+0.40}_{-0.03}$) and \citetalias{KimYJ+2020} ($\alpha \sim -1.2^{+1.36}_{-0.64}$), which are based on a deep and wide-area surveys ($> 80$ deg$^2$), within 1-$\sigma$ level. \citet{McGreer+2018} derived a steeper faint-end slope ($\alpha \sim -1.97^{+0.09}_{-0.09}$) than ours. However, their value should be taken with a caution since the LF of \citet{McGreer+2018} is based mostly on quasars with $M_{1450} < -24$ mag. \citet{Kulkarni+2019} also showed a steep $\alpha$ of $\sim -2.31$ using the same dataset of \citet{McGreer+2013} plus \citet{Glikman+2011}, but their data points are limited to $M_{1450} < -24$ \citep{McGreer+2018} or a small number statistics due to a coverage of $\sim 2$ deg$^{2}$ \citep{Glikman+2011}. Clearly, this comparison demonstrates how uncertain the LF faint-end slope could be without sufficiently deep data.

\indent \citet{Giallongo+2019} and \citet{Boutsia+2018} presented a near infrared (NIR) + X-ray selected AGN UV LF, but their LF at faint end is about 10 $\times$ higher than ours. \citet{Shen+2020} discuss a possible tension of AGN LF of \citet{Giallongo+2019} with other LFs. The AGN LF of \citet{Giallongo+2019} has been considered as a possible evidence for quasars making non-negligible contribution to IGM ionization at $z \sim 5$ (e.g., \citealp{Grazian+2020, Grazian+2022}). Our LF, along with other previous LFs, shows a rather low number density of faint AGNs, supporting claims for quasars contributing little in IGM ionization at $z \sim 5$ (\citealp{McGreer+2018}; \citetalias{KimYJ+2020}; \citealp{Shin+2020}). 

\indent \citetalias{Niida+2020} used color-selected quasar candidates as well as a few spectroscopically confirmed quasars to derive quasar LF. They excluded the binned LFs at $M_{1450} \lesssim -23.3$ mag due to possible contamination by Lyman break galaxies (LBGs). On the other hand, we extend our LF to $M_{\rm 1450} \lesssim -22.0$ mag, over one mag fainter than the \citetalias{Niida+2020} limit. At $M_{\rm 1450} = -22.0$ mag, the quasar number density from our work is several times smaller than \citetalias{Niida+2020} LF, suggesting an efficient rejection of LBGs through our selection method. We discuss this point in detail in the next subsections.

\indent Finally, we note that our binned LF at $-22 < M_{1450} < -21$ is comparable to the LFs of \citet{Giallongo+2019} and \citet{Boutsia+2018}, although our faintest quasar sample is significantly contaminated by LBGs. The expected level of contamination is very high, with $\gtrsim 12 \%$ of LBGs contaminating the quasar sample at these magnitudes (Section~\ref{sec:dis2}). Correcting such a level of LBG contamination would bring the binned LF points down to the extrapolated portion of the parametric LF or below it. Therefore, the binned LFs at $-22 < M_{1450} < -21$ mag can serve as another evidence against $z\sim5$ quasars making a significant contribution to the IGM ionization. One caveat is that AGNs at these magnitudes may not appear as quasars (point-like sources) and have their light dominated by host galaxies. Our selection method would miss such objects since we pick up point sources with quasar-type SEDs as quasars (e.g., \citealt{KimYJ+2021}).

\begin{figure*} 
\centering 
\includegraphics[width=0.9\textwidth]{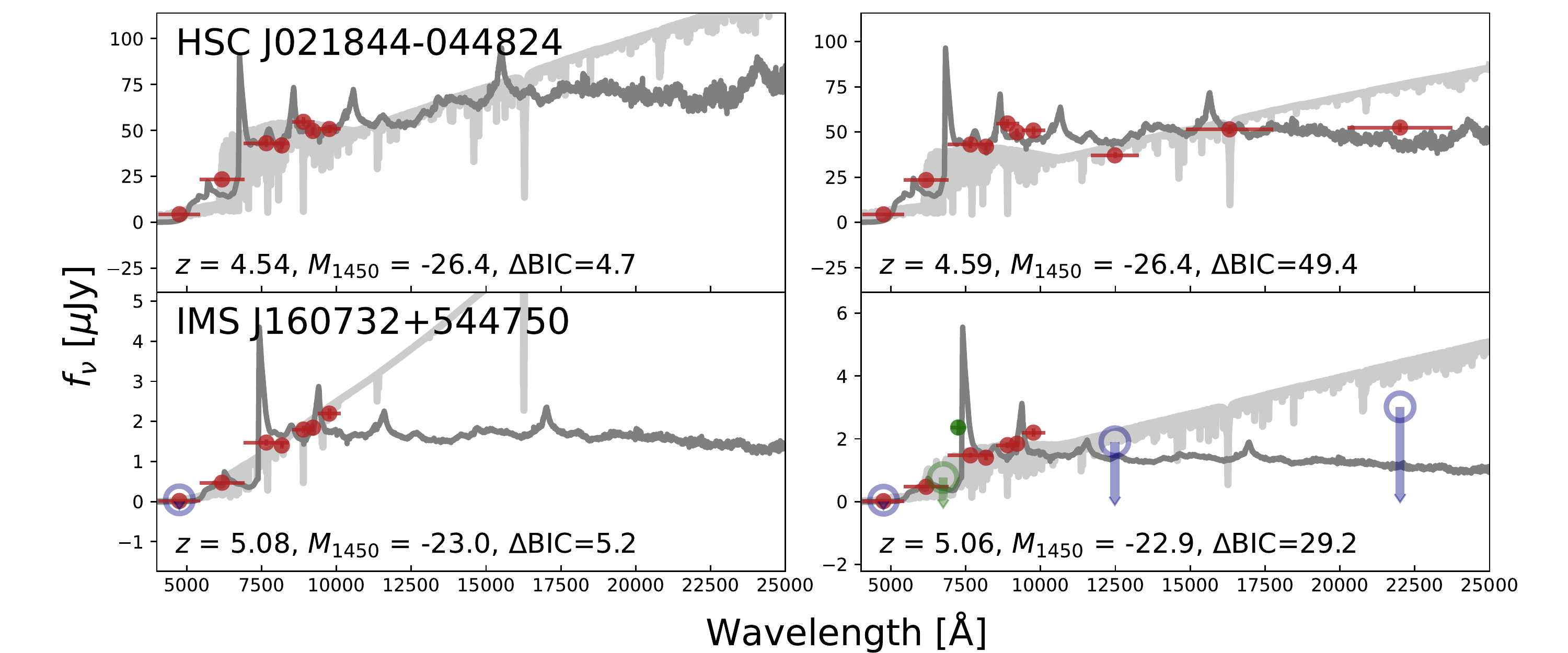}
\caption{The SED-fitting results of one quasar (HSC J021844-044824) and one medium-band selected quasar (IMS J160732+544750) in \citet{Shin+2020}. For a given object, we show the results with the seven HSC-SSP filters (left) and ten filters at least (right) consisting of HSC-SSP, near-infrared, and medium-band filters in \citet{Shin+2020}. The dark gray and the light gray lines show the best-fit quasar SED model and star SED model, respectively. The measured fluxes by HSC-SSP and near-infrared filters (red), and medium-band filters (green) are marked with points. The blue open circles with arrows indicate the detection limit in the case of non-detection.}
\label{fig:6_Recovery}
\end{figure*}

\subsection{Recovery of known quasars with our selection} \label{sec:dis2}
We examine how many known quasars are recovered by our selection method. There are six $z\sim5$ quasars identified by spectroscopy in our survey area \citep{McGreer+2013, Paris+2018, Shin+2020} and three medium-band selected quasars reported in \citet{Shin+2020}\footnote{medium-band selected quasars indicate quasar candidates with $\Delta$BIC $\gtrsim 30$ that are highly likely to be real quasars at $z\sim5$ based on multiwavelength measurement from UV to NIR.}. While five of the 6 spectroscopically identified quasars and two of the medium-band selected quasars are recovered as the final sample, HSC J021844-044824 (spectroscopically confirmed) and IMS J160732+544750 (a medium-band selected quasar in \citealt{Shin+2020}) dissatisfy the $\Delta$BIC criterion with $\Delta$BIC $\sim5$ in this work. Missing 2 out of 9 quasars (22 $\%$ of the sample) can be explained by the completeness of the survey. The brighter one, HSC J021844-044824 is at $z=4.5$ which corresponds to the parameter space where the completeness is low ($\sim 0.27$). IMS J160732+544750 at $z\sim5$ has a low luminosity of $M_{1450}=-22.9$ mag where the completeness starts declining rapidly.

\indent As demonstrated in the right panels of Figure~\ref{fig:6_Recovery}, the selection can be improved with additional filters. For example, IMS J160732+544750 was selected as a quasar with an addition of a medium-band and NIR upper limits \citep{Shin+2020}. HSC J021844-044824 could have been selected as a quasar if there was an additional NIR band data. In conclusion, our selection method with HSC-SSP photometry data may miss $\sim 22\%$ (2/9) of known quasars at $z\sim5$. However, the quasar recovery can be improved by including additional multi-band data.

\begin{figure*} 
\includegraphics[width=0.95\textwidth]{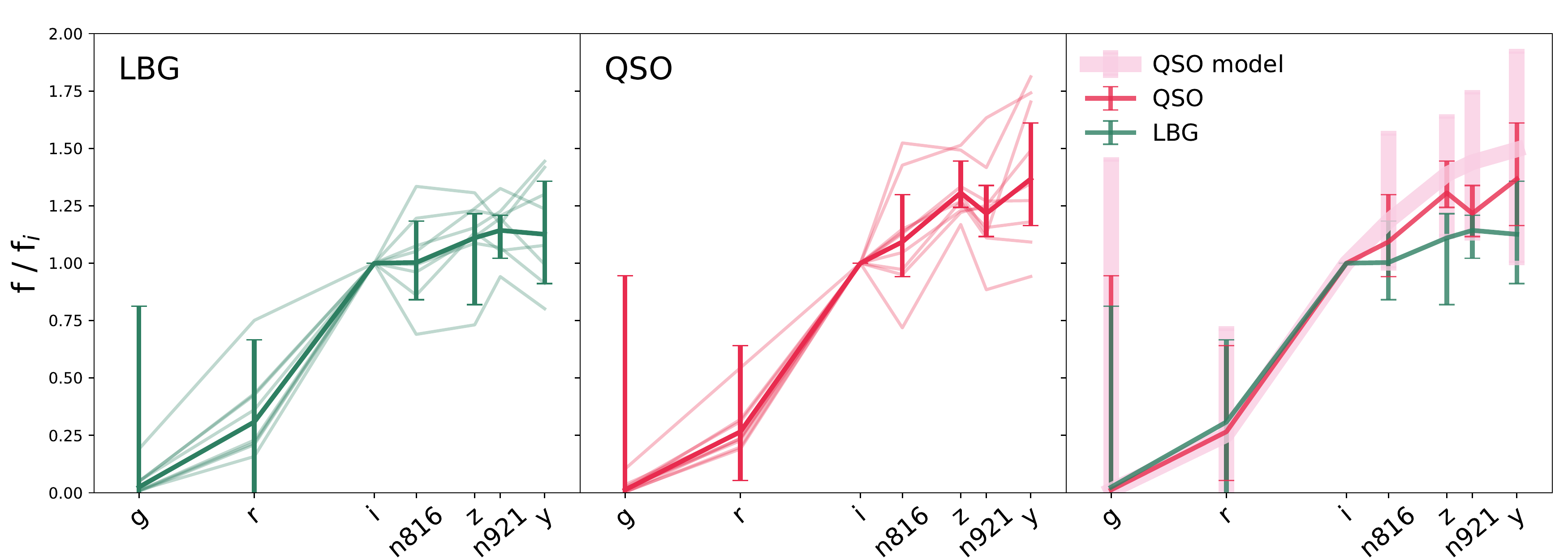}
\centering 
\caption{The $i$-band normalized SEDs of the galaxies at $z\sim5$ (‘LBG’) and quasars/the promising candidates at $z\sim5$ (‘QSO’). The light red and green solid lines in the left and middle panels show the SED of each object in the ‘QSO’ and ‘LBG’ samples, respectively. The red and green solid lines correspond to the mean fluxes of the ‘QSO’ and ‘LBG’ objects. The pink thick line in the right panel represents the mean fluxes of model quasars at $z\sim5$. The error bars with the solid lines indicate a 68 $\%$ range of normalized fluxes.}
\label{fig:7_ContRate}
\end{figure*}

\subsection{Contamination rate of our quasar survey}
\subsubsection{High-redshift galaxies} \label{sec:dis1}
We checked if our quasar sample is contaminated by high-redshift galaxies using spectroscopically confirmed galaxies at $z \sim 4-7$ from HSC-SSP \citep{Ono+2018}. This is because of two reasons: 1) Several quasar surveys have indicated a high contamination rate of high-redshift quasars samples by high-redshift galaxies in a faint regime (\citealt{Matsuoka+2018_2}; \citetalias{Niida+2020}). The number density of LBGs is significantly higher than that of quasars at $M_{1450} > -23$ mag (\citealp{Ono+2018}; see Figure 9), and 2) our selection has no explicit criteria for separating high-redshift galaxies from quasars at z $\sim 5$. Specifically, the DL selection considers the $i$-band-selected point sources only, and the BIC statistics makes use of the star models only. 

\indent We searched for spectroscopically confirmed galaxies at $z\sim5$ in our survey area and found 8 galaxies with $i < 25$ mag at $4.0 < z_{\rm spec} < 6.0$. Among them, we find only one galaxy at z $\sim 5.0$ satisfy the pre-, DL, and $\Delta$BIC criteria of our quasar selection, meaning that the contamination of the sample by Lyman-break galaxies is low at $12.5 \%$ (1/8). The pre-selection reduced the galaxy number from 8 to 5. The DL selection removes another galaxy, and the $\Delta$BIC calculation removes 3 out of 4 galaxies passing the pre- and DL selections.

\indent We compared the galaxy contamination rate of our selection method with a traditional high-redshift quasar selection made from a color-color diagram. High-redshift objects show a distinctive feature in color-color space due to redshifted Lyman break, and are often selected from a specific region in a color-color space (e.g., see \citealt{Shim+2007}; \citealt{Kang+2009} for galaxies; \citealt{Choi+2012}; \citealt{KimYJ+2015}; \citealt{Jeon+2017} for quasars). \citetalias{Niida+2020} selected $z\sim5$ quasars based on their broad-band colors and point source appearance, and noticed possible, significant contamination of their sample by galaxies at $M_{1450} \sim -23$ mag. We applied the \citetalias{Niida+2020} color selection criteria to our point sources, finding that the galaxy contamination fraction is 3$/$8 (37.5 $\%$ contamination rate). This example demonstrates that our quasar selection adopting the DL and BIC statistics can lower the contamination rate of the quasar sample by compact galaxies by a few times. 

\indent The role of each selection process in selecting final quasar candidates and evaluating the suitability of its criterion are discussed further below. First, we checked the dependency of the contamination rate on the point source selection cut in the pre-selection process. If a looser cut ($i-i_{\rm CModel}<$ 0.3) was adopted instead of 0.2, we have three more galaxies (i.e., $8/8$ of known galaxies in the area) as quasar candidates in the pre-selection process. However, after applying the DL and BIC selections, the remaining galaxy is one at $z\sim5.0$, which is identical to the case of the tighter point source cut. The contamination rate of the quasar selection process is insensitive to the choice of point source selection if the cut is realistic enough to contain almost all of the point sources.

\indent DL selects $4/5$ galaxies satisfying the pre-selection criteria. This means that our DL process is not very effective in excluding $z\sim5$ galaxies from the sample. On the other hand, the $\Delta$BIC calculation based on the best-fit result of SED-fitting could exclude $3/4$ remaining galaxies in quasar candidates, suggesting its effectiveness in reducing the galaxy contamination.

\indent To figure out the reason why the $\Delta$BIC calculation improves the faint quasar selection, we compared the SEDs of 9 quasar/quasar candidates \citep{McGreer+2018, Shin+2020} with those of 8 LBGs \citep{Ono+2018} with $i<25$ at $4.0 < z_{\rm spec} < 6.0$ by normalizing them to their $i$-band fluxes. Then, we calculated the mean SED for the quasar (‘QSO’ class) and LBGs (‘LBG’ class), and the 68th percentile region of two classes for each band. In Figure~\ref{fig:7_ContRate}, we show the difference between the mean SEDs of the two classes. The mean SED of the ‘QSO’ sample has redder colors than that of the ‘LBG’ sample due to the bluer continuum slope of LBGs than quasars at high redshift (e.g., \citealt{Jiang+2013}). This distinctive feature could not be adequately sampled in color selections using 3 to 4 broadband filters, whereas our selection could extract the feature and filter out candidates whose SEDs are dissimilar to the quasar SED models.

\indent Since the BIC is related to the $\chi^{2}$ value of the best-fit model, the absolute value of $\Delta$BIC decreases as the photometric uncertainties increase. Indeed, the fractions of the DL-selected candidates satisfying the $\Delta$BIC$>10$ at $i<23$ mag and $i>23$ mag are $\sim$ 0.19 and 0.05, respectively. Therefore, the exclusion of galaxies from the quasar sample through the $\Delta$BIC selection may be merely due to the $\Delta$BIC selection preferentially excluding faint objects ($i<23$ mag) with large photometric errors. Hence, we estimated how the photometric uncertainties influence the $\Delta$BIC selection at fainter magnitudes.

\indent To do so, we added median magnitude errors of the HSC-SSP sources with $i=24$ mag to DL-selected candidates with $i<23$ mag, and repeated the $\Delta$BIC selection. Note that $i=24$ mag is close to the $i$-band magnitude of the faintest quasar candidate. The fraction of DL-selected candidates passing the $\Delta$BIC selection decreases only moderately from the original 0.19 to $\sim 0.14$ for this noise-added sample, not as much as 0.05 in the real data. Therefore, the photometric error only partially explains the decreasing fraction of the excluded candidates through the $\Delta$BIC selection at fainter magnitudes. This result supports our suggestion that the $\Delta$BIC selection excludes high-redshift galaxies efficiently, which are expected to be more numerous among the fainter UV high-redshift sources. Additional deep spectroscopy of the final candidates would validate this conclusion.

\subsubsection{Stars and quiescent galaxies}
Common contaminants in high-redshift quasar surveys are faint stars whose colors are similar to those of quasars (e.g., \citealt{Matsuoka+2019_10}). To estimate the fraction of stars satisfying our selection criteria, we generated a mock catalog of 100,000 stars uniformly distributed in $i < 24$ based on stellar model spectra (Section~\ref{sec:stellar_model}) and scaled their SEDs to $i=24$ mag. Their magnitude errors are assigned by randomly selecting a value from a Gaussian distribution of which standard deviations are median magnitude errors of HSC-SSP sources that have similar magnitudes to theirs. We adopted this approach because few spectroscopically confirmed faint stars ($i>22.5$ mag) are in the Deep layer of the HSC-SSP. Among the 100,000 randomly sampled mock stars, 100,000 and 20,131 stars pass the pre-selection criteria (1) and (3), assuming the extendedness values of all stars are within 0.2. 3,213/20,131 stars are DL-selected candidates, however, the BIC selection finds no promising candidate in the 3,213 candidates due to their redder colors ($i-NB816$, $i-z$, $i-NB921$, and $i-y$) than those of quasars. As a result, the fraction of pre-selected faint stars passing our DL and $\Delta$BIC selection criteria is 0/20,131.

\indent Another possible contaminants are quiescent galaxies at $0.5 < z < 1.0$ whose 4,000 \AA~ breaks can mimic sharp Lyman breaks of quasars at $z\sim5$ \citep{Euclid+2019}. To test how many quiescent galaxies can be selected with our selection process, we prepared a catalog of quiescent galaxies at $0.5 < z < 1.0$ from \citet{Weaver+2022} in the COSMOS field, which are selected to be objects that form red envelope in the $r-i$ color and photometric redshift space (e.g., \citealt{Im+2002}). In a 0.3 deg$^2$ area where the multi-band HSC-SSP Deep and COSMOS fields overlap, we identified 1,847 quiescent galaxies with $i\leq24$ mag at $0.5 < z < 1.0$ within a matching radius of $1\farcs0$.

\indent These galaxies have a median $i$-band magnitude of $\sim 22$ mag. Hence, we rescaled their SEDs to $i=24$ mag by adding the difference between $i=24$ and their $i$-band magnitudes to their SEDs. We increased the quiescent galaxy sample size up to 100,000 by assigning photometric uncertainties to the rescaled magnitudes. The uncertainties were randomly given in the same way as the mock stars. Among the 100,000 simulated galaxies, 87,801 galaxies meet the pre-selection criteria (1) and (3) based on an assumption that they satisfy our point source selection cut. Although DL classifies 538/87,801 galaxies as qso candidates, the $\Delta$BIC calculation filters out all the candidates, resulting in the fraction of pre-selected quiescent galaxies satisfying our DL and $\Delta$BIC selection of 0/87,801.

\indent Given that there are 125,644 sources passing the pre-selection criteria, the very low contamination rates by pre-selected, simulated stars (0/20,131) and quiescent galaxies at $0.5 < z < 1.0$ (0/87,801) imply that these two types of objects cannot significantly contaminate the quasar candidates sample. This is because that photometric errors of HSC-SSP Deep data are small enough at our effective depth of $i<24$ to accurately trace red colors of late-type stars and detect $g$-band fluxes of quiescent galaxies at good SNRs. When we calculated the FPR of the DL and BIC selection for nqso objects, it was as low as 0.015$\%$. This expected FPR is in line with the fractions calculated from the above results.


\begin{figure*} 
\includegraphics[width=\textwidth]{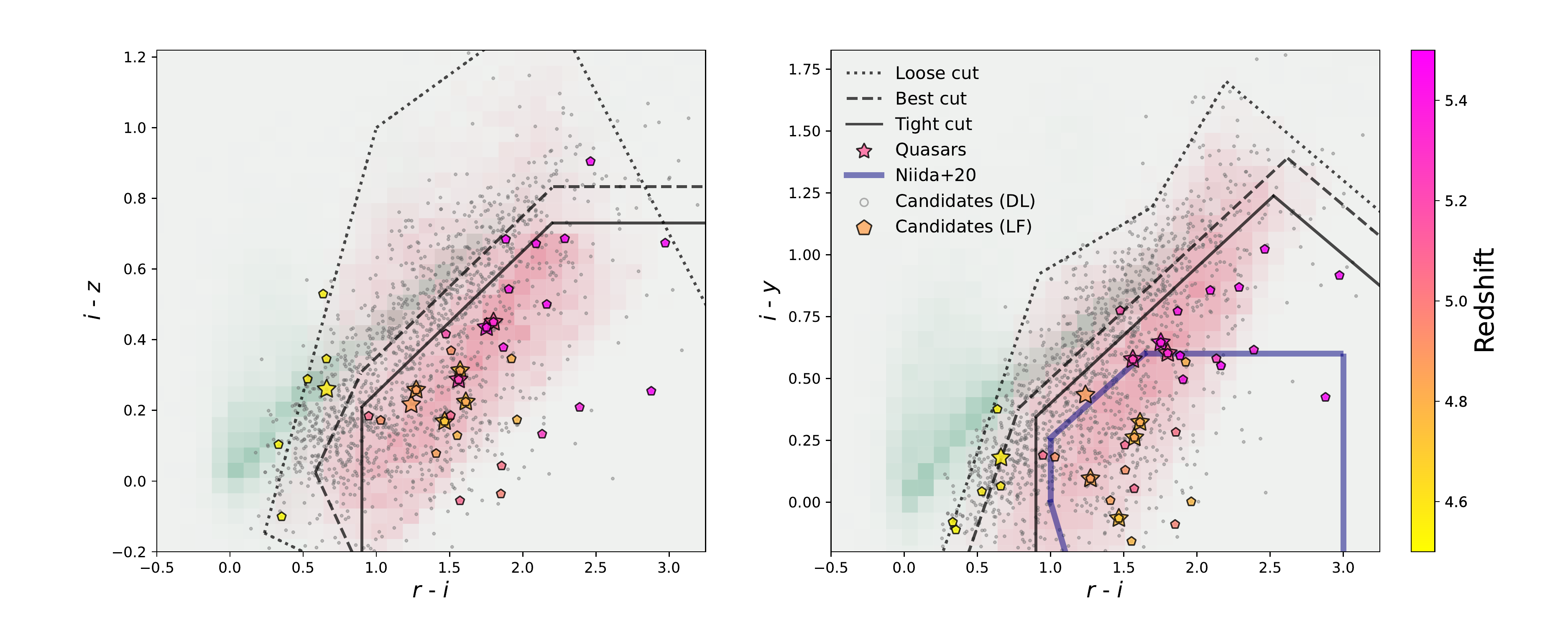}
\centering 
\caption{Color-color diagrams in $riz$ (left) and $riy$ (right). The green and red histograms indicate the number densities of $i$-band-selected point sources with $i<25$ mag and quasar models. The 1,599 DL-selected candidates are marked with gray dots. The stars indicate the confirmed quasar or promising candidates reported in \citet{Shin+2020}, while the pentagons represent the 35 quasar candidates used for constructing the parametric LF. These stars and pentagons are color-coded by their photometric redshifts. The dotted, dashed, and solid lines correspond to the ‘Loose', ‘Best', and ‘Tight' color selection criteria, respectively. Color cuts used in \citetalias{Niida+2020} are marked with the navy lines in the $riy$ color space.} 
\label{fig:8_CCD}
\end{figure*}

\begin{figure*} 
\includegraphics[width=\textwidth]{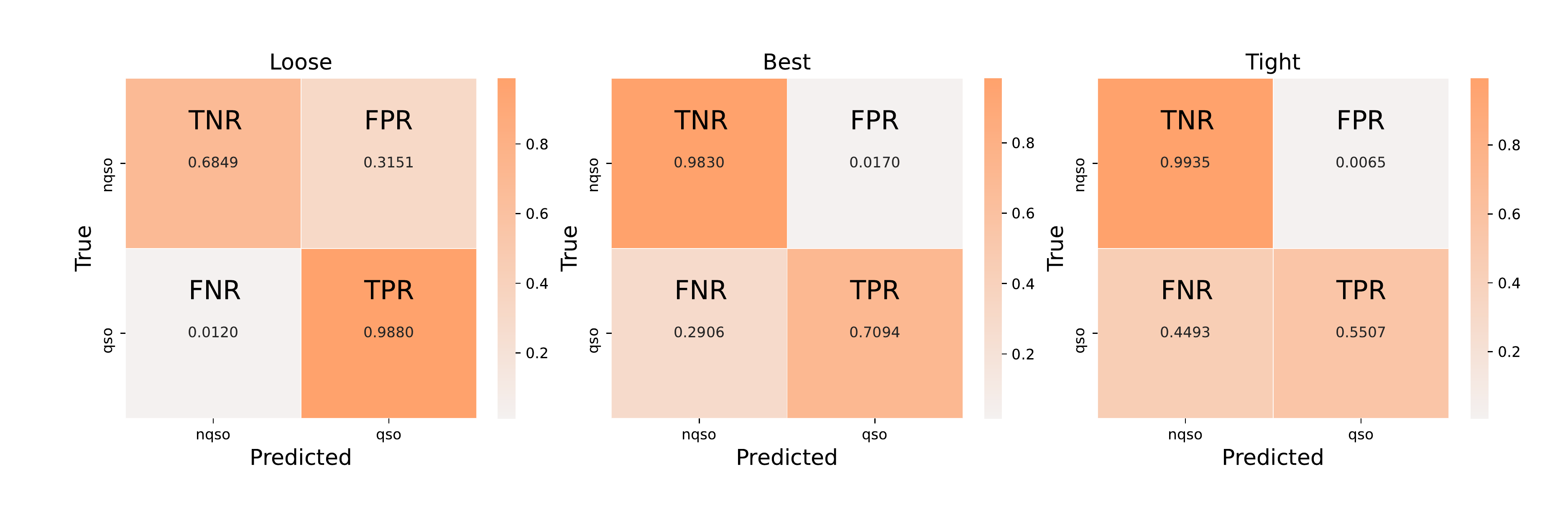}
\centering 
\caption{The confusion matrices of different color selection criteria, ‘Loose', ‘Best', and ‘Tight', calculated in the same way as we compute for the confusion matrix of deep learning shown in Figure~\ref{fig:2_ConfMat}.}
\label{fig:9_ConfMat_CC}
\end{figure*}

\subsection{Comparison of DL and color selections} \label{sec:comp_cc_vs_DL}
To compare the performance of the DL selection with that of traditional color selection method, we devised three different types of color cuts (‘Loose', ‘Best', ‘Tight') in $r-i$ versus $i-z$ ($riz$) and $r-i$ versus $i-y$ ($riy$) spaces. These cuts are guided by considering the distribution of quasar models at $z=4.5-5.5$ on each color space. The ‘Loose' cut represents a selection cut for maximizing the recovery rate of the quasar models (i.e., minimizing the miss rate of quasars, minimizing FNR), whereas the ‘Tight' cut indicates a selection cut for minimizing contamination rate (i.e., minimizing FPR). The ‘Best' cut stands for an optimal cut selecting as many quasars as possible while minimizing contaminants. Figure~\ref{fig:8_CCD} shows these cuts, our DL-selected candidates, and 35 final candidates. As shown in the figure, even if the ‘Loose' cut is applied, two of the 35 candidates are excluded, implying that DL selection is effective in including more quasars in the candidate sample. The ‘Tight' cut encloses 74 $\%$ of the 35 candidates, whereas the color cuts used in \citetalias{Niida+2020} miss about a half of the candidates, since the cuts of \citetalias{Niida+2020} are tailored for quasars only at $z=4.7$ to $5.1$.

\indent Figure~\ref{fig:9_ConfMat_CC}, we present the confusion matrices for the three cases of color criteria. The ‘Loose' cut has the highest recovery rate of quasar models (e.g., True Positive Rate, TPR) among the three cases, but the contamination rate in the quasar selection is also very high ($\sim 30 \%$). The ‘Tight' cut shows a very low contamination rate in the quasar sample, but misses $\sim 45 \%$ of quasar models. The ‘Best' cut has relatively reasonable FPR and TPR, however, its FPR is $\sim 3$ times larger than that of DL selection ($\sim 0.5 \%$) and its TPR is well below that of the DL selection ($\sim 100 \%$). Compared to color selections, the DL selection gives a high recovery rate of quasars with a low FPR. For example, the DL method can select quasar candidates at $z\sim4.5$ which is difficult to do so in the color cut method. One can loosen the color cuts to make it as inclusive as DL for quasar selection. But, this makes FPR too large ($60 \times$ or more of DL), and hence selects too many contaminants that the $\Delta$BIC selection needs to weed out (1,500 for DL versus 9,000 for color cut). DL is a more complete, efficient selection method than traditional color selections. A shortcoming of DL is that it cannot account for the difference between the absolute number of nqso and qso objects, but this can be augmented with an additional selection procedure such as the $\Delta$BIC calculation.



\bigskip
\section{Summary} \label{sec:Summary}
\indent To construct a quasar LF at $z\sim5$, we selected quasars using a new technique that combines the DL and $\Delta$BIC selections. Quasars were chosen from the Deep layer of the HSC-SSP imaging survey covering 15.5 $\mathrm{deg}^{2}$.

\indent We found that our selection outperforms traditional selection methods based on color cuts, sampling more quasars at a wider redshift range while minimizing contamination from LBGs. The former advantage was made possible by DL with its flexible color criteria. The latter merit was achieved through the $\Delta$BIC selection that enabled us to distinguish quasar SEDs from bluer SEDs of LBGs. Compared to the color selection of \citetalias{Niida+2020}, we achieved three times less contamination rate by galaxies at $z\sim5$. Our selection process recovered most of confirmed quasars as well.

\indent Thanks to our selection, we constructed a $z\sim5$ quasar LF reaching $M_{1450} = -22.0$ mag, about 1 magnitude deeper than previous LFs. The overall shape of the LF is similar to LFs in recent works at $z\sim5$ (\citetalias{Niida+2020}; \citetalias{KimYJ+2020}) down to $M_{1450} \sim -24.0$ mag, indicating a flatter faint-end slope of $\alpha = -1.60^{+0.21}_{-0.19}$ than some previous studies \citep{McGreer+2018, Kulkarni+2019}. We even tried to estimate the LF at -22 to -21 mag. Knowing that the faintest quasar sample could be contaminated by LBGs significantly (about 10$\%$ or more of LBGs classified as quasars), the LF at the faintest bins agrees with the flatter faint-end slope. These results suggest that quasars -- AGNs with point-like appearance -- are not contributing significantly to the IGM ionization.

\indent In this paper, we demonstrated the feasibility of our selection and the importance of attempting a novel and efficient approach to select promising quasar candidates from numerous faint objects. Future spectroscopic observations of our final quasar candidates will confirm the validity of our method, and adding multi-wavelength data would help select high-redshift quasars more reliably.

\acknowledgments
We appreciate the anonymous referee's useful comments. We thank Jinyi Yang for providing the quasar sample and completeness function of the bright quasar survey. This research was supported by the National Research Foundation of Korea (NRF) Grant (No. 2020R1A2C3011091 and No. 2021M3F7A1084525), funded by the Ministry of Science and ICT (MSIT). S. S. acknowledges the support from the Basic Science Research Program through the NRF funded by the Ministry of Education (No. 2020R1A6A3A13069198). Y. K. was supported by the NRF grant funded by the MSIT (No. 2021R1C1C2091550) and acknowledges the support from the China Postdoc Science General (2020M670022) and Special (2020T130018) Grants funded by the China Postdoctoral Science Foundation.

\indent The Hyper Suprime-Cam (HSC) collaboration includes the astronomical communities of Japan and Taiwan, and Princeton University. The HSC instrumentation and software were developed by the National Astronomical Observatory of Japan (NAOJ), the Kavli Institute for the Physics and Mathematics of the Universe (Kavli IPMU), the University of Tokyo, the High Energy Accelerator Research Organization (KEK), the Academia Sinica Institute for Astronomy and Astrophysics in Taiwan (ASIAA), and Princeton University. Funding was contributed by the FIRST program from the Japanese Cabinet Office, the Ministry of Education, Culture, Sports, Science and Technology (MEXT), the Japan Society for the Promotion of Science (JSPS), Japan Science and Technology Agency (JST), the Toray Science Foundation, NAOJ, Kavli IPMU, KEK, ASIAA, and Princeton University. 

\indent This paper makes use of software developed for the Large Synoptic Survey Telescope. We thank the LSST Project for making their code available as free software at  http://dm.lsst.org

\indent This paper is based on data collected at the Subaru Telescope and retrieved from the HSC data archive system, which is operated by Subaru Telescope and Astronomy Data Center (ADC) at National Astronomical Observatory of Japan. Data analysis was in part carried out with the cooperation of Center for Computational Astrophysics (CfCA), National Astronomical Observatory of Japan.

\indent This research has made use of the NASA/IPAC Extragalactic Database (NED), which is funded by the National Aeronautics and Space Administration and operated by the California Institute of Technology.

\vspace{5mm}
\software{RayTune \citep{Liaw+2018}, scikit-learn \citep{Pedregosa+2011}, astroquery \citep{astroquery+19}, emcee \citep{emcee+2013}}

\newpage
\bibliography{ref}{}
\bibliographystyle{aasjournal}

\end{document}